\documentstyle[11pt,aasms4]{article}

\received{October 30 1998}
\accepted{May 7 1999}

\begin{document}

\title{The $L_x$-$T$ and $L_x$-$\sigma$ Relationships for Galaxy Clusters 
       Revisited} 

\author{Xiang-Ping Wu and Yan-Jie Xue}

\affil{National Astronomical Observatories, Chinese Academy
                 of Sciences, Beijing 100012, China}
\and
\author{Li-Zhi Fang}

\affil{Department of Physics, University of Arizona, Tucson, AZ 85721}

\begin{abstract}
The relationships between the X-ray determined bolometric luminosity $L_x$,  
the temperature $T$ of the intracluster gas, and the optical measured 
velocity dispersion $\sigma$ of the cluster galaxies are updated 
for galaxy clusters using the largest sample of 256 clusters
drawn from literature. The newly established relationships, based on 
the doubly weighted orthogonal distance regression (ODR) method, 
are justified by both their self-consistency 
and co-consistency, which can then be used to
test the theoretical models of cluster formation and evolution.
The observationally determined $L_x$-$T$ and $L_x$-$\sigma$ relationships,
$L_x\propto T^{2.72\pm0.05}\propto \sigma^{5.24\pm0.29}$,
are marginally consistent with those predicted in the scenario that
both intracluster gas and galaxies are in isothermal and hydrostatic 
equilibrium with the underlying gravitational potential of clusters.
A comparison between these observed and predicted $L_x$-$T$ relationships 
also suggests that the mean cluster baryon fraction $f_b$ remains 
approximately constant among different clusters, 
$f_b\approx0.17$, which gives rise to
a low-mass density universe of $\Omega_m\approx0.3$.
\end{abstract}


\section{Introduction}

Over the past few years, there has been an increase in the observational
and theoretical studies of the relationships between the X-ray determined 
bolometric luminosity $L_x$, 
the temperature $T$ of the intracluster gas 
and the optical measured velocity dispersion $\sigma$ of the cluster galaxies
(e.g. Edge \& Stewart 1991a,b; David et al. 1993; 
Fabian et al. 1994; Girardi et al. 1996; 
Mushotzky \& Scharf 1997; Cavaliere, Menci \& Tozzi 1997;
Scharf \& Mushotzky 1997; White, Jones \& Forman 1997;  
Markevitch 1998; Wu, Fang \& Xu 1998; 
Arnaud \& Evrard 1998; Allen \& Fabian 1998;  etc.). 
On one hand, because the total X-ray luminosity of a cluster is related to 
the (gas) baryon number density through thermal bremsstrahlung,
while the X-ray temperature and galaxy velocity dispersion are determined 
by the gravitating mass of the cluster, the $L_x$-$T$ and $L_x$-$\sigma$ 
relationships constitute a direct and sensitive probe of
the volume-averaged baryon fraction $f_b$ of the cluster.  
On the other hand, because both gas and galaxies are the tracers of 
the depth and  shape of a common gravitational potential in the cluster, 
the correlation between $T$ (or $L_x$) and $\sigma$ provides a crucial test 
for the dynamical properties of galaxy clusters.
Moreover, the possible intrinsic dispersions
in the $L_x$-$T$, $L_x$-$\sigma$ and $\sigma$-$T$ relationships
are likely to be associated with the cluster merging histories, 
preheating of intracluster gas, 
the presence of cooling flows or systematic variations in
the baryon fraction with cluster masses. Furthermore, 
the observed $L_x$-$T$, $L_x$-$\sigma$ and $\sigma$-$T$  
relationships may also allow one to constrain the models of structure 
formation (e.g. Evrard \& Henry 1991; 
Lubin et al. 1996; Mushotzky \& Scharf 1997; 
Donahue et al. 1998; Eke, Navarro \& Frenk 1998; etc).

It has been known since the early days of X-ray astronomy that there
exists a strong correlation between the following dynamical properties
of clusters: $L_x$ vs $T$ (Mitchell, Ives \& Culhane 1977; see Table 2 for
a summary),  $L_x$ vs $\sigma$ (Solinger \& Tucker 1972) 
and $T$ vs $\sigma$ [Smith, Mushotzky 
\& Serlemitsos 1979; For a summary see Table 2 of Wu, Fang \& Xu (1998)].  
While the accuracies in determinations
of these relationships have been significantly improved in recent
years with the rapid growth of X-ray/optical data for galaxy clusters, 
tight constraints on these relationships have not yet been satisfactorily 
achieved. The main reason is that there may exist intrinsic dispersions in
these relationships due to different physical mechanisms among 
different clusters. Another reason is the small cluster samples
involved in statistics.  Yet, there is also a reason related to  
the inappropriate linear regression methods employed in the fitting process.  
For the latter a typical example is the fitted $\sigma$-$T$ 
relationship for galaxy
clusters: The early claim for $\sigma\propto T^{0.50}$, 
which implies a perfect isothermal and hydrostatic equilibrium for 
the dynamical state of galaxy clusters (Lubin \& Bahcall 1993),
has been recently shown to depend on the adopted 
ordinary standard least-square (OLS) fitting method (Wu, Fang \& Xu 1998  
and references therein). An apparent deviation of the dynamical
state from that predicted by the isothermal and hydrostatic equilibrium
model for galaxy clusters is found if the doubly weighted orthogonal 
distance regression (ODR) is applied for the same data set ($\sigma$, $T$).
Indeed, some of the previous fitted relationships based on 
a sparse data set,  together with the inappropriate linear regression 
methods,  could be misleading. The first purpose of this paper is then to
demonstrate the uncertainties in the $L_x$-$T$ and $L_x$-$\sigma$  
relationships for galaxy clusters
arising from the employment of the OLS and ODR fitting methods. 
To this end, 
we will use the hitherto largest cluster sample drawn from literature,
which enables us to update the cluster $L_x$-$T$ and $L_x$-$\sigma$
relationships and remarkably increase the statistical significance.
The reliability of the new relationships will be tested by both 
self-consistency and co-consistency between different correlations.
Next, we will attempt to derive the mean cluster baryon fraction $f_b$ 
and the average mass density of the Universe
from the newly established relationships between $L_x$, $T$ and $\sigma$.
Furthermore, we will briefly discuss the possibility of
testing various theoretical models and speculations with 
our  $L_x$-$T$ and  $L_x$-$\sigma$ relationships.
Throughout the paper we assume a Hubble constant of 
$H_0=50$ km s$^{-1}$ Mpc$^{-1}$ and a flat cosmological
model of $\Omega_0=\Omega_m=1$.

\section{Sample}

By extensively searching the literature, we find 168/193 galaxy clusters 
for which both the X-ray bolometric luminosity and temperature/velocity 
dispersion are observationally determined (Table 1). 
Here, we have excluded those clusters whose $L_x$, $T$ or $\sigma$ are
derived from indirect methods such as the $L_x$-$T$, $L_x$-$\sigma$
or $\sigma$-$T$ correlations.  While this leads to a reduction of cluster 
numbers as compared with the largest cluster sample heretofore published
by White et al. (1997), we have added many high-redshift and newly 
discovered X-ray clusters. 
Since the cluster X-ray luminosities are measured in different energy bands, 
we convert the observed luminosities to the bolometric luminosities 
in the rest frame of the clusters 
according to an optically thin and isothermal plasma emission 
model, in which we adopted the analytic approximation of the total
Gaunt factor given by Mewe et al. (1986) and assumed
$T=6$ keV if the X-ray temperatures are unknown.
For some clusters whose X-ray luminosities are originally given within 
a certain cluster radius, we have made a
correction for lost flux falling out the detection apertures by
the standard method which assumes a $\beta$ model with 
$\beta=2/3$ and $r_c=0.25$ Mpc for the X-ray surface brightness distribution. 
The final sample contains a total of 256 clusters which constitutes 
the hitherto largest cluster data set for such studies.


\section{The $L_x$-$T$ relationship}

Bolometric luminosities $L_x$ against temperatures $T$ 
for 168 clusters in Table 1 are plotted in Fig.1,
in which we have also explicitly illustrated
the distributions of the high-redshift ($z>0.1$) and low-redshift ($z>0.1$) 
clusters. While the scatter is large, there is a strong correlation between 
the two variables. Employing the OLS to the whole data set yields 
\begin{equation}
L_x=10^{-0.92\pm0.09}T^{2.61\pm0.12},
\end{equation}
where (also hereafter) $L_x$ and $T$ are in units of $10^{44}$ erg s$^{-1}$ 
and keV, respectively. Since measurement uncertainties remain unknown in
some cases, they are not included in the error bars in the above fitting.
An immediate impression is 
that the resulting $L_x$-$T$ relationship is roughly consistent with 
previous analyses (see Table 2), in particular the recent result by 
Markevitch (1998) where the central cooling regions
are excluded. Additionally, no apparent difference in the $L_x$-$T$ 
distributions is seen  among the low- and high-redshift clusters. 
Next, we extract from Table 1 a subsample of 142 clusters that have 
measurement uncertainties in both quantities $L_x$ and $T$ (see Fig.2). 
We apply respectively the OLS and ODR methods to the subsample, which gives
\begin{eqnarray}
L_x=10^{-0.86\pm0.10}T^{2.52\pm0.13}, & ({\rm OLS});\\
L_x=10^{-0.92\pm0.05}T^{2.72\pm0.05}, & ({\rm ODR}).
\end{eqnarray}
The error bars in these fits are determined by Monte-Carlo simulations,
which have taken the measurement uncertainties in both $L_x$ and $T$ into
account. 

\placefigure{fig1}
\placefigure{fig2}

Although the discrepancy in the fitted $L_x$-$T$ relationships by
different authors has partially arisen from the selection of cluster
samples such as whether the cooling clusters are included/excluded 
(White et al. 1997; Markevitch 1998; Allen \& Fabian 1998), another
reason is apparently due to the adopted linear regression methods:
The ODR fitting method usually provides a steeper slope than that given 
by the OLS method in the fitted relationships. A similar problem is also
reported in the study of the $\sigma$-$T$ relationship (Wu, Fang \& Xu 1998). 
Because OLS ignores the scatters in $T$ and only minimizes the residuals 
in $L_x$ while ODR makes an attempt at accounting for scatters 
in both $T$ and $L_x$ (Feigelson \& Babu 1992),  in principle
ODR provides a more reasonable fit than OLS for our subsample
in the sense that  $L_x$ and especially $T$ contain significant measurement 
uncertainties in addition to their intrinsic scatters (also 
see White et al. 1997).
This can be seen by a self-consistent test: Using $L_x$ as the
abscissa variable and repeating the above fitting process, we obtain
\begin{eqnarray}
T=10^{0.45\pm0.01}L_x^{0.28\pm0.01},  & (168\; {\rm points}, {\rm OLS}); \\
T=10^{0.44\pm0.02}L_x^{0.29\pm0.01}, &  ({\rm 142\; points}, 
                                        {\rm OLS});\\
T=10^{0.34\pm0.01}L_x^{0.37\pm0.01}, &  ({\rm 142\; points}, 
                                         {\rm ODR}).
\end{eqnarray}
If $L_x=10^aT^b$, we would expect that $T=10^{-a/b}L_x^{1/b}$. 
It is immediate that the ODR fitted relationships meet this
simple criteria while the OLS results, $L_x\propto T^{2.61}$ (or
$T^{2.52}$) and $T\propto L_x^{0.28}$ (or $L_x^{0.29}$), 
are inconsistent with each other.
In summary, based on our cluster sample and the ODR method, we find
that at the $3\sigma$ level, $L_x\propto T^{2.6}$ -- $T^{2.9}$ and
$T\propto L_x^{0.34}$ -- $L_x^{0.40}$.


\section{The $L_x$-$\sigma$ relationship}

Similarly, the relationship between X-ray bolometric luminosity
and galaxy velocity dispersion can be established for a total of
193 clusters in Table 1 and a subsample of 156 clusters whose
measurement uncertainties in both $L_x$ and $\sigma$ are known,
respectively,
\begin{eqnarray}
L_x=10^{-7.06\pm0.54}\sigma^{2.67\pm0.19},  & (193\; {\rm points}, 
                                             {\rm OLS}); \\
L_x=10^{-6.73\pm0.59}\sigma^{2.56\pm0.21}, &  ({\rm 156\; points}, 
                                        {\rm OLS});\\
L_x=10^{-14.57\pm0.94}\sigma^{5.24\pm0.29}, &  ({\rm 156\; points}, 
                                         {\rm ODR}),
\end{eqnarray}
or
\begin{eqnarray}
\sigma=10^{2.76\pm0.01}L_x^{0.19\pm0.01},  & (193\; {\rm points}, 
                                             {\rm OLS}); \\
\sigma=10^{2.75\pm0.01}L_x^{0.20\pm0.02}, &  ({\rm 156\; points}, 
                                        {\rm OLS});\\
\sigma=10^{2.78\pm0.03}L_x^{0.19\pm0.01}, &  ({\rm 156\; points}, 
                                         {\rm ODR}),
\end{eqnarray}
where (also hereafter)  $\sigma$ is in units of km s$^{-1}$. 
The data sets ($L_x,\sigma$), together with our best fitted results, 
are shown in Fig.3 and Fig.4 for 193 and 156 clusters, respectively.
Self-consistent test suggests that at the $3\sigma$ level,
$L_x\propto \sigma^{4.4-6.1}$, which is compatible with 
the recent work of White et al. (1997) based on 50 clusters and the ODR 
fitting technique. Recall that
the previous studies in terms of the OLS fitting method found that
$L_x\propto \sigma^{4}$ (Quintana \& Melnick 1982) or
$L_x\propto \sigma^{2.9}$ (Edge \& Stewart 1991b).

\placefigure{fig3}
\placefigure{fig4}

We have further plotted in Fig.5  the $\sigma$-$T$ relationship 
for 105 clusters in Table 1. Applying the ODR fitting technique 
to the 92 clusters, for which measurement uncertainties in both $T$ and
$\sigma$ are given, yields 
\begin{equation}
\sigma=10^{2.49\pm0.03}T^{0.64\pm0.02}.
\end{equation}
This is in good agreement with that given by Wu, Fang \& Xu (1998) 
for a sample of 95 clusters where the calculated  
X-ray temperatures for some clusters from White et al. (1997) were used.  
On the other hand, our fitted correlations 
between $L_x$ and $T$/$\sigma$, 
$L_x=10^{-0.92\pm0.05}T^{2.72\pm0.05}$ and
$L_x=10^{-14.57\pm0.94}\sigma^{5.24\pm0.29}$, provide 
\begin{equation}
\sigma=10^{2.60\pm0.24}T^{0.52\pm0.07}.
\end{equation}
It appears that within a $2\sigma$ uncertainty range, 
the `derived' $\sigma$-$T$ relationship is consistent with 
the observed one [eq.(13)].
This is the first time that   
the co-consistency between these correlations has been established.

\placefigure{fig5}

\section{Cosmic evolution}

Different theoretical models predict quite different cosmic evolutionary 
tendencies for galaxy clusters, 
and thus an examination of whether the observed X-ray 
luminosities, temperatures and velocity dispersions of galaxy clusters evolve 
with redshift may discriminate among these theoretical models. For instance,  
in the scenario of self-similar hierarchical structure 
formation of galaxy clusters,
the evolution of the cluster population follows  
$L_x\propto (1+z)^{(7n+5)/(2n+6)}$ and $T\propto\sigma^2
\propto (1+z)^{(n-1)/(n+3)}$,
where $n$ is the spectral index of the density perturbation,
$P(k)\propto k^n$ (Kaiser 1986). 
Generally, clusters undergo rapid evolution in such a model.
So, by comparing with X-ray observations, in principle we can test this
scenario and thus determine the power index $n$. Unfortunately, 
a stringent constraint on the evolutionary models of clusters cannot be set
because our X-ray cluster sample  (Table 1) is neither from a complete
flux-limited survey nor from a temperature-limited one.
Nevertheless, the redshift dependence of X-ray luminosities, 
temperatures and velocity dispersions for all the clusters 
in Table 1 (see Fig.6) may give 
useful implications about cluster evolution. 
It appears that 
while the scatters are large, evidence for the cosmic evolution of 
these dynamical quantities is rather weak, which is consistent with the result
obtained from studies of cluster temperature evolution (Henry, Jiao
\& Gioia 1994; Mushotzky \& Scharf 1997) and 
X-ray luminosity function as well as distribution of X-ray core radii 
of clusters
(e.g. Burke et al. 1997; Rosati et al. 1998; Vikhlinin et al. 1998).

\placefigure{fig6}

\section{The mean cluster baryon fraction}

For simplicity we assume that the hot X-ray emitting gas in a cluster 
follows a spherically symmetrical isothermal $\beta$ model with $\beta=2/3$.
The electron number density is then
\begin{equation}
n_e=n_{e0}\left(1+\frac{r^2}{r_{xc}^2}\right)^{-1},
\end{equation}
where $n_{e0}$ and $r_{xc}$ are the central electron number density and
the core radius, respectively. 
In the scenario of the optically thin and isothermal plasma emission,
the X-ray bolometric luminosity within a cluster radius $r$ is given by
\begin{equation}
L_x=\frac{2^4e^6}{3\hbar m_ec^2}\left(\frac{2\pi kT}{3m_ec^2}\right)^{1/2}
     \mu_e \overline{g}(T) \int n_e^24\pi r^2dr,
\end{equation}
where $\overline{g}(T)$ is the average Gaunt factor, and $\mu_e=2/(1+X)$ with
$X=0.768$ is the
hydrogen mass fraction in the primordial abundances of hydrogen and helium.
The gas mass within the same volume is 
\begin{equation}
M_{gas}=\mu_e m_p\int n_e 4\pi r^2 dr.
\end{equation}
The total gravitating mass of a cluster $M_{tot}$ can be independently derived 
using the hydrostatic equilibrium equation for X-ray emitting gas
and galaxies, respectively,
\begin{eqnarray}
-\frac{GM_{tot}}{r^2}=\frac{kT}{\mu_im_p}\frac{d n_e}{n_e dr},\\
-\frac{GM_{tot}}{r^2}=\sigma^2\frac{d n_{gal}}{n_{gal} dr},
\end{eqnarray}
where $\mu_i=0.585$ denotes the mean molecular weight, $n_{gal}$ is
the number density of cluster galaxies for which we use
\begin{equation}
n_{gal}=n_{gal,0}\left(1+\frac{r^2}{r_{gc}^2}\right)^{-\alpha},
\end{equation}
$n_{gal,0}$ and $r_{gc}$ are the central galaxy number density
and core radius, respectively, and $\alpha$ is the power-law index
which has a value of $\alpha=3/2$ and $1$ for the conventional King model
and the softened isothermal model, respectively.  Defining the
(gas) baryon fraction as $f_b=M_{gas}/M_{tot}$, we can write the
X-ray bolometric luminosity $L_x$ as\\
{\noindent}(a)for gas as the tracer of the cluster potential 
\begin{equation}
L_x=\frac{2^4e^6}{3\hbar m_ec^2}\left(\frac{2\pi kT}{3m_ec^2}\right)^{1/2}
     \frac{\overline{g}(T)}{\mu_e\mu_i^2}
     \left(\frac{kT}{Gm_p^2}\right)^2f_b^2S_{gas},
\end{equation}
where the structure factor $S_{gas}$ is
\begin{equation}
S_{gas}=\frac{1}{2\pi r_{xc}}\left(\frac{r^2}{r^2+r_{xc}^2}\right)^2
         \frac{\tan^{-1} \frac{r}{r_{xc}}-\frac{rr_{xc}}{r^2+r_{xc}^2} }
	      {\left(1-\frac{r_{xc}}{r}\tan^{-1} \frac{r}{r_{xc}}\right)^2 }; 
\end{equation}
{\noindent}(b)for galaxies as the tracer of the cluster potential 
\begin{equation}
L_x=\frac{2^4e^6}{3\hbar m_ec^2}\left(\frac{2\pi kT}{3m_ec^2}\right)^{1/2}
     \frac{\overline{g}(T)}{\mu_e}
     \left(\frac{\sigma^2}{Gm_p}\right)^2f_b^2S_{gal},
\end{equation}
where the structure factor $S_{gal}$ is
\begin{equation}
S_{gal}=\frac{\alpha}{2\pi r_{xc}}\left(\frac{r^2}{r^2+r_{gc}^2}\right)^2
         \frac{\tan^{-1} \frac{r}{r_{xc}}-\frac{rr_{xc}}{r^2+r_{xc}^2} }
	      {\left(1-\frac{r_{xc}}{r}\tan^{-1} \frac{r}{r_{xc}}\right)^2 }. 
\end{equation}
In the computation of total X-ray luminosity of the whole cluster,
we have $r\gg r_{xc}$ and $r\gg r_{gc}$, which gives
$S_{gas}\approx 1/4r_{xc}$ and $S_{gal}\approx \alpha/4r_{xc}$.
In addition, the mean Gaunt factor depends only 
weakly on gas temperature. Therefore,  the X-ray bolometric luminosity
scales approximately as\\
{\noindent}(a)for gas as the tracer of the cluster potential 
\begin{equation}
L_x\propto T^{2.5} f_b^2 r_{xc}^{-1};
\end{equation}
{\noindent}(b)for galaxies as the tracer of the cluster potential 
\begin{equation}
L_x\propto \sigma^{4}T^{1/2} f_b^2 r_{xc}^{-1}
\propto T^{3.06} f_b^2 r_{xc}^{-1},
\end{equation}
in which the $\sigma$-$T$ relationship [eq.(13)] has been used.

The fact that the theoretically expected results [eq.(25) and eq.(26)] are 
marginally consistent with the
observationally fitted $L_x$-$T$ relationship [eq.(3)]  
indicates that $f_br_{xc}^{-1}$ remains roughly unchanged 
for different clusters. The universality of $f_b$ for
different clusters is naturally expected 
since a typical rich cluster draws its matter 
(baryon + non-baryon) from a region of radius of $\sim20$ Mpc which should
be large enough to be representative of the matter composition of the 
Universe. On the other hand, there has been no strong evidence so far for the
dependence of the X-ray core radii on temperature (Edge \& Stewart 1991a), 
although a recent study reported a weak trend
of increasing core radius with increasing gas temperature
(Jones \& Forman 1999). In particular, 
an essentially identical distribution of 
core radii of nearby and distant luminous clusters has been detected
(Vikhlinin et al. 1998), showing that $r_{xc}$ does not evolve significantly 
with redshift since $z\approx1$.  Therefore, our finding that 
$f_br_{xc}^{-1}$ holds statistically constant for clusters within $z\approx1$ 
is reasonable.

\subsection{X-ray emitting gas as the tracer of cluster potential}

If cluster matter composition is representative of the universe, 
the volumed-average (gas) baryon fractions over the whole cluster 
should remain the same for different clusters, i.e., $f_b=const$.
Assuming that the gas fraction $f_b$ 
and core radius $r_{xc}$ are constant for different clusters, and 
the intracluster gas is in  isothermal and hydrostatic equilibrium with
the gravitational potential of cluster,
we would expect that $L_x\propto T^{2.5}$ in terms of eq.(25), which
has a somewhat flatter slope than that observed [eq.(3)]  
unless the OLS fitted $L_x$-$T$ relationships eqs.(1) and (2) are invoked.
We plot in Fig.7(b) the ratio of $L_x$ to $T^{2.5}$ for the 168 clusters 
whose $L_x$ and $T$ are observationally determined. 
Regardless of the large scatter,  
the ratio $L_x/T^{2.5}$ is nearly independent of 
cluster population, with a mean value of 
$\langle L_x/T^{2.5}\rangle=0.098\pm0.091$. 
Because there is an apparent asymmetry in the distribution of
the $L_x/T^{2.5}$ ratios,  we provide 
the median and $90\%$ limits of the distribution: 
$\langle L_x/T^{2.5}\rangle =0.135^{+0.295}_{-0.105}$. This gives rise to 
the mean cluster gas fraction 
\begin{equation}
\begin{array}{ll}
f_b & =0.195 \left(\frac{L/T^{2.5}}{0.135}\right)^{1/2}
    \left(\frac{r_{xc}}{0.25\;{\rm Mpc}}\right)^{1/2}\\
    &          \\
    & =0.145^{+0.153+0.081}_{-0.103-0.072},
\end{array}
\end{equation}
in which the first uncertainty account for the $90\%$ limits of 
the $L_x/T^{2.5}$ distribution and the second one arises from 
the variation of core radius from 0.1 Mpc to 0.5 Mpc. 
We have used a value of $\overline{g}(T)=1.2$ for the mean Gaunt factor, which
introduces another uncertainty of about $20\%$.

\placefigure{fig7}

In the self-similar model for the formation of galaxy clusters,
if the X-ray emission is dominated by bremsstrahlung, 
the expected $L_x$-$T$ relationship eq.(25) reduces to 
\begin{equation}
L_x\propto T^{2} (1+z)^{3/2},
\end{equation}
where clusters are assumed to have  same $f_b$ and formed at same redshift.
In order to test the self-similar models of cluster evolution,
we plot the ratio of 
$L_x$ to $T^2(1+z)^{3/2}$ in Fig.7(c) 
for our cluster sample (Table 1). It appears that the dispersions are 
rather large as compared with the similar plot for the $L_x/T^{2.5}$
ratio. Essentially, there is no correlation between
$L_x$ and $T^2(1+z)^{3/2}$.
The flatter slope of the $L_x$-$T$ relationship ($L_x\propto T^2$) and 
the possible $L_x$ evolution with redshift (depending on $n$) 
in the self-similar models are apparently contrary to observations.
To reconcile with observations, a number of physical mechanisms have 
been explored such as preheating of the intracluster gas 
(Ponman, Cannon \& Navarro 1999 and references therein)
or the influence of cooling flows (Allen \& Fabian 1998). 
Here, we discuss briefly another possibility
suggested by David et al. (1993):  
One can achieve the observed relationship 
$L_x\propto T^{3}$ by simply requiring
the baryon fraction to vary as $f_b\propto T^{0.5}$.  Although the recent
work by Allen \& Fabian (1998) arrived at an opposite conclusion that
$f_b\propto T^{-0.68\pm0.22}$, their gas mass fractions $f_b$ are measured
at a relatively small radius of $0.5$ Mpc, 
which may have rather large uncertainties,
in addition to the intrinsic dispersion, 
due to the problem of conventional cluster mass estimates and 
dynamical activities in the central regions (Wu \& Fang 1997; 
Allen 1998; Wu et al. 1998 and references therein). 
Indeed, the binding mass of a cluster within
the central core could be underestimated by a factor of $2$--$4$ 
due to the employment of the isothermal and hydrostatic equilibrium 
hypothesis [eq.(18)]. So, 
the baryon fraction can be overestimated by a corresponding
factor within the cluster core. Nevertheless, the various mass estimators
(X-ray, optical and gravitational lensing)  
can provide a consistent cluster mass on scale greater than the X-ray
core. This may explain the deceasing
tendency of $f_b$ with $T$ reported by Allen \& Fabian (1998) if we notice that
$T\propto r^{2}$ in a self-similar model of cluster growth
(e.g. Evrard, Metzler \& Navarro 1996). Namely, the cluster baryon
fraction might be more significantly overestimated for the low-temperature 
clusters than the high-temperature ones. 
However, this speculation needs to be tested
by a combined analysis of cluster mass estimates from
gravitational lensing and X-ray measurements. 
In summary, if the baryon fraction of a cluster 
is dependent on temperature, it may have significant impact on our
determination of the universal baryon fraction based on a flux limited cluster 
sample.

\subsection{Galaxies as the tracer of cluster potential}

Unlike the diffuse X-ray gas, cluster galaxies are less affected by the
presence of the (non)cooling flows, the nonthermal pressure 
(e.g. magnetic field), preheating (e.g. supernovae), etc. 
Therefore, there are good reasons that cluster galaxies are a better tracer 
of the  underlying gravitational potential of clusters 
and their velocity dispersions 
$\sigma$ are a good indicator of dark matter. 
The convincing evidence comes from the excellent agreement
between the gravitating masses of clusters derived from the gravitational 
lensing phenomena and from the velocity dispersion of galaxies as 
a tracer of cluster potential (Wu \& Fang 1997; Wu et al. 1998).  
Nevertheless,  the expected $L_x$-$T$ relationship [eq.(26)] 
under the assumption that galaxies are in hydrostatic equilibrium with 
the cluster gravitational potential shows a slightly steeper slope than 
that observed [eq.(3)]. The gas fraction $f_b$ according to  eq.(23) is
\begin{equation}
f_b^2  =3.725\left(\frac{L_x/T^{3.06}}{0.1}\right)
	\left(\frac{\sigma/T^{0.64}}{100}\right)^{-4}
	\left(\frac{r_{xc}}{0.25{\rm Mpc}}\right)\frac{1}{\alpha}.
\end{equation}
By fixing the power-law index $3.06$ for the $L_x$-$T$ relationship
and $0.64$ for the $\sigma$-$T$ relationship in the fitting process, 
we have  
\begin{equation}
L_x=10^{-1.14\pm0.03}T^{3.06},
\end{equation}
and 
\begin{equation}
\sigma=10^{2.49\pm0.02}T^{0.64}.
\end{equation}
The mean cluster baryon fraction is 
\begin{equation}
f_b=0.172\pm0.006\pm0.016\pm0.071\pm0.034\pm0.013
\end{equation}
where the first and second error bars account for the uncertainties in 
the fitted $L_x$-$T$ and $\sigma$-$T$ relationships, respectively, 
the third and forth ones 
correspond to the variations of core radius $r_{xc}$ and $\alpha$ 
in the ranges
$0.1$ Mpc $\le r_{xc}\le$ $0.5$ Mpc and $0.7<\alpha<1.5$, 
and the last term is the uncertainty introduced by the mean Gaunt factor.
Our statistical estimates of cluster baryon fraction, both eq.(27) and 
eq.(32), are consistent
with the results obtained for many individual clusters based on 
detailed analysis of the dynamical properties of clusters 
(e.g. White et al. 1993; David, Jones \& Forman 1995; 
Wu 1998).

\subsection{Cosmological density parameter}

In conjunction with the standard Big Bang Nucleosynthesis (BBN) model 
(Walker et al. 1991),  we can estimate the cosmological density 
parameter $\Omega_m$ from the baryon fractions given above
(e.g. White et al. 1993; David et al. 1995). 
(a)For gas as the tracer of the cluster potential,
\begin{equation}
\Omega_m=0.30\pm0.06\pm0.11\pm0.09       , 
\end{equation}
in which the first error bar accounts for 
the uncertainty of the BBN prediction: $\Delta\Omega_b=0.01$ where 
$\Omega_b$ is the average baryon mass density of the Universe,
the second one represents the uncertainty in the fitted $L_x$-$T$ relationship
and the adopted mean Gaunt factor, and the third term reflects
the variation of core radius $r_{xc}$ from $r_{xc}=0.1$ Mpc to 
$r_{xc}=0.5$ Mpc. (b)For galaxies as the tracer of the cluster potential
\begin{equation}
\Omega_m=0.29\pm0.06\pm0.06\pm0.46, 
\end{equation}
where the first error bar accounts again for 
the uncertainty of the BBN prediction, and the
second one is the combined result of
uncertainties in the fitted $L_x$-$T$ and $\sigma$-$T$ relationships
and the mean Gaunt factor, and the last one includes the uncertainties 
due to luminous matter distributions ($r_{xc}$ and $\alpha$).
Strictly speaking, the above estimate of $\Omega_m$ is in 
contradiction with our cosmological model assumed at the first onset:
$\Omega_0=\Omega_m=1$. We have then re-estimated the X-ray luminosity
$L_x$ for each cluster in our subsample by converting the currently 
used X-ray luminosity in an $\Omega_0=\Omega_m=1$ model to that 
in a flat cosmological model but with a nonzero cosmological constant, 
$\Omega_m=0.3$ and $\lambda=0.7$. 
Nevertheless, we find that this change has 
a negligible effect on our $L_x$-$T$, $L_x$-$\sigma$ and $\sigma$-$T$ 
relationships.

\section{Discussion and conclusions}

Regardless of possible intrinsic dispersions, overall clusters exhibit
a strong correlation between the X-ray luminosity, the temperature of
the hot intracluster gas and the velocity dispersion of cluster galaxies.
Based on the largest cluster sample drawn from literature and 
the ODR fitting method, that takes the measurement uncertainties of 
two variables into account,  we have established both
self-consistent and co-consistent relationships between these quantities.
These relationships may  enable us to test theoretical
models of cluster formation and evolution and determine the mean baryon
fraction and the average mass density of the Universe.

A comparison between the theoretically expected 
$L_x$-$T$ and $L_x$-$\sigma$ relationships for clusters and
the observationally determined ones
suggests that both the intracluster gas and 
galaxies are marginally in isothermal and hydrostatic equilibrium 
with the underlying gravitational potential of clusters, which is 
consistent with the result given by a combined study of cluster
mass determinations from the X-ray/optical and gravitational lensing 
measurements (Allen 1998; Wu et al. 1998). Therefore, it is unlikely that
the local dynamical activities can play a dominant role in the 
recent dynamical evolution of overall clusters. 
However, neither the X-ray emitting gas nor the cluster 
galaxies as a tracer of the gravitational potential can precisely reproduce 
the observed $L_x$-$T$ relationships. Moreover, our 
fitted $\sigma$-$T$ relationship, $\sigma\propto T^{0.64}$, 
shows an apparent deviation from that predicted by isothermal 
and hydrostatic model for galaxy clusters,  $\sigma\propto T^{0.5}$.
This may have arisen from our oversimplification in modeling the  
gas/galaxy distributions. 

We have shown that the mean cluster (gas) baryon fraction is  
roughly constant among different clusters with a mean value of
$f_b\approx0.17$, which corresponds to  
$\Omega_m\approx0.3$. Indeed, the volume-averaged 
baryon fractions of different clusters over the whole cluster sizes 
should be the same, if clusters are a fair sample of the Universe.
Our estimated cluster baryon fraction and 
the average matter density of the Universe are in accordance with 
the prevailing claim for a low-mass density universe with or without 
a nonzero cosmological constant.

We have not found convincing evidence for a significant evolution in the  
dynamical properties of clusters characterized by
$L_x$, $T$ and $\sigma$, although the highest redshift cluster at $z\approx1$
has been included in our cluster sample. This disagrees 
with the claims by the standard CDM model and the self-similar models
for formation and evolution of clusters, but is consistent with 
the results from numerous
recent X-ray observations (e.g. Mushotzky \& Scharf 1997;
Rosati et al. 1998; etc.) and the predictions by 
the low-mass density cosmological models.

Apparently,  the present conclusions are subject to the incompleteness 
of our cluster sample, which is neither flux-limited nor 
temperature-limited (or velocity dispersion-limited). 
It is likely that there would be  a moderate 
modification to our $L_x$-$T$, $L_x$-$\sigma$ and $\sigma$-$T$
relationships when a large and complete cluster sample becomes 
available. In particular, the large scatters in these correlations
are likely to be associated with the intrinsic 
properties of clusters, especially
the cooling flows in the central regions (Fabian et al. 1994; 
White et al. 1997; Markevitch 1998; Allen \& Fabian 1998), 
which have not been separated in our fitting processes.  
Finally, future investigations should be made towards 
a deep understanding of the physical mechanisms for the reported
$L_x$-$T$, $L_x$-$T$ and $\sigma$-$T$ relationships.

\acknowledgments
We gratefully acknowledge the insightful comments and suggestions by
an anonymous referee. This work was supported by 
the National Science Foundation of China, under Grant No. 19725311.

\clearpage


\setcounter{page}{23}

 \begin{deluxetable}{llllll}
 \tablewidth{30pc}
 \scriptsize
 \tablecaption{Cluster Sample}
 \tablehead{
 \colhead{cluster }& \colhead{redshift} &
 \colhead{$\sigma$(km/s)} & 
 \colhead{$T$ (keV)} &  
 \colhead{$L_x$ ($10^{44}$ erg/s)} & 
 \colhead{references}  }
 \startdata
 A13            &0.0905&$  896^{+  85}_{-  73}$& & $   5.03^{+  1.17}_{-  1.17}$&F96 $\cdot\cdot\cdot$ E96 \nl
 A21            &0.0948&   621 & & $    8.12 ^{+  1.59}_{-  1.59}$&S+R $\cdot\cdot\cdot$ E96 \nl
 A76            &0.0416& & $ 1.50^{+ 1.10}_{- 0.60}$&$   1.01^{+  0.22}_{-  0.22}$&$\cdot\cdot\cdot$ D93 E96 \nl
 A85            &0.0559&$  810^{+  76}_{-  80}$&$ 6.20^{+ 0.40}_{- 0.50}$&$  19.52^{+  1.35}_{-  1.35}$&B95 DJF E96 \nl
 A115           &0.1971&  1167 & & $   31.09 ^{+  7.42}_{-  7.42}$&S+R $\cdot\cdot\cdot$ E96 \nl
 A119           &0.0438&$  863^{+ 178}_{- 112}$&$ 5.59^{+ 0.27}_{- 0.27}$&$   7.17^{+  0.82}_{-  0.82}$&ZHG HMS E96 \nl
 A133           &0.0604&$  735^{+  87}_{-  72}$&$ 4.00^{+ 1.40}_{- 0.90}$&$   6.67^{+  0.75}_{-  0.75}$&WQI D93 E96 \nl
 A151           &0.0537&$  708^{+  69}_{-  55}$& & $   2.31^{+  0.39}_{-  0.39}$&GEF $\cdot\cdot\cdot$ E96 \nl
 A154           &0.0652&$  843^{+ 276}_{- 142}$& & $   1.83^{+  0.10}_{-  0.10}$&DDD $\cdot\cdot\cdot$ J+F \nl
 A160           &0.0447&   572 & & $    0.87 ^{+  0.28}_{-  0.23}$&S+R $\cdot\cdot\cdot$ E98 \nl
 A168           &0.0438&$  435^{+  31}_{-  24}$&$ 2.60^{+ 1.10}_{- 0.60}$&$   1.50^{+  0.25}_{-  0.25}$&F96 D93 E98 \nl
 A189           &0.0335&$  259^{+  94}_{-  49}$& & $   0.21^{+  0.05}_{-  0.05}$&ZHG $\cdot\cdot\cdot$ E98 \nl
 A193           &0.0490&$  723^{+  78}_{-  61}$&$ 4.20^{+ 1.00}_{- 0.60}$&$   3.05^{+  0.60}_{-  0.60}$&F96 D93 E98 \nl
 A194           &0.0184&$  341^{+  57}_{-  37}$&$ 2.63^{+ 0.15}_{- 0.15}$&$   0.22^{+  0.04}_{-  0.04}$&F96 HMS E96 \nl
 A222           &0.2110&   570 & & $    7.65 ^{+  0.33}_{-  0.33}$&S+R $\cdot\cdot\cdot$ DFJ \nl
 A262           &0.0169&$  525^{+  47}_{-  33}$&$ 2.41^{+ 0.05}_{- 0.05}$&$   0.86^{+  0.08}_{-  0.08}$&F96 A+E E98 \nl
 A272           &0.0872&$  694^{+ 193}_{- 115}$& & $   7.36^{+  1.55}_{-  1.55}$&ZHG $\cdot\cdot\cdot$ E96 \nl
 A370           &0.3730&$ 1340^{+ 230}_{- 150}$&$ 7.13^{+ 1.05}_{- 1.05}$&$  20.77^{+  1.95}_{-  1.95}$&FMB M+S FMB \nl
 A376           &0.0489& & $ 5.00^{+ 2.00}_{- 1.10}$&$   2.88^{+  0.42}_{-  0.42}$&$\cdot\cdot\cdot$ D93 E98 \nl
 A399           &0.0718&$  961^{+  71}_{-  55}$&$ 7.40^{+ 0.50}_{- 0.50}$&$  78.99^{+  9.63}_{-  9.63}$&GEF A+E E98 \nl
 A400           &0.0237&$  599^{+  80}_{-  65}$&$ 2.31^{+ 0.14}_{- 0.14}$&$   0.60^{+  0.12}_{-  0.12}$&F96 HMS E96 \nl
 A401           &0.0737&$ 1152^{+  86}_{-  70}$&$ 8.00^{+ 0.40}_{- 0.40}$&$  26.42^{+  2.45}_{-  2.45}$&F96 HMS E98 \nl
 A407           &0.0472&   597 & & $    1.15 ^{+  0.30}_{-  0.30}$&S+R $\cdot\cdot\cdot$ E98 \nl
 A426           &0.0179&$ 1277^{+  95}_{-  78}$&$ 6.79^{+ 0.12}_{- 0.12}$&$  31.75^{+  0.15}_{-  0.15}$&ZHG HMS E96 \nl
 A458           &0.1054&$  736^{+  86}_{-  58}$& & $   6.00^{+  1.20}_{-  1.20}$&F96 $\cdot\cdot\cdot$ E96 \nl
 A478           &0.0881&$  904^{+ 261}_{- 140}$&$ 6.90^{+ 0.35}_{- 0.35}$&$  32.00^{+  4.08}_{-  4.08}$&ZHG HMS E98 \nl
 A483           &0.2830&&$ 8.70^{+ 2.00}_{- 1.30}$&   48.90  & $\cdot\cdot\cdot$ D93 D93 \nl
 A496           &0.0325&$  687^{+  89}_{-  76}$&$ 4.13^{+ 0.08}_{- 0.08}$&$   6.81^{+  0.60}_{-  0.60}$&F96 HMS D99 \nl
 A514           &0.0714&$  882^{+  84}_{-  64}$& & $   3.27^{+  0.64}_{-  0.64}$&F96 $\cdot\cdot\cdot$ E96 \nl
 A520           &0.2010&$  988^{+  76}_{-  76}$&$ 8.59^{+ 0.93}_{- 0.93}$&$  37.35^{+  8.41}_{-  8.41}$&C96 M+S E98 \nl
 A539           &0.0284&$  832^{+  77}_{-  60}$&$ 3.24^{+ 0.09}_{- 0.09}$&$   1.85^{+  0.21}_{-  0.21}$&ZHG HMS E96 \nl
 A545           &0.1530& & $ 5.50^{+ 6.20}_{- 1.10}$&$  21.66^{+  2.49}_{-  2.49}$&$\cdot\cdot\cdot$ D93 E96 \nl
 A548           &0.0416&$  853^{+  62}_{-  51}$& & $   2.63^{+  0.79}_{-  0.79}$&ZHG $\cdot\cdot\cdot$ D99 \nl
 A548S          &0.0415&&$ 2.40^{+ 0.70}_{- 0.50}$&    1.30  & $\cdot\cdot\cdot$ D93 D93 \nl
 A569           &0.0201&$  327^{+  95}_{-  39}$& & $   0.30^{+  0.06}_{-  0.06}$&F96 $\cdot\cdot\cdot$ A+K \nl
 A576           &0.0384&$  945^{+  93}_{-  88}$&$ 4.30^{+ 0.30}_{- 0.30}$&$   2.71^{+  0.43}_{-  0.43}$&F96 D93 E98 \nl
 A578           &0.0864&  793&$ 1.70^{+ 1.50}_{- 0.60}$&$   5.97^{+  0.48}_{-  0.48}$&G97 G97 G97 \nl
 A586           &0.1710& & $ 6.61^{+ 1.15}_{- 0.96}$&$  25.27^{+  4.95}_{-  4.95}$&$\cdot\cdot\cdot$ M+S E98 \nl
 A644           &0.0704& & $ 6.59^{+ 0.17}_{- 0.17}$&$  18.92^{+  2.17}_{-  2.17}$&$\cdot\cdot\cdot$ A+E E96 \nl
 A665           &0.1816& 1201&$ 8.26^{+ 0.90}_{- 0.90}$&$  41.72^{+  6.51}_{-  6.51}$&S+R A+E E98 \nl
 A671           &0.0501&   994 & & $    2.07 ^{+  0.44}_{-  0.44}$&S+R $\cdot\cdot\cdot$ E98 \nl
 A744           &0.0732&$  814^{+ 173}_{- 106}$& & $   1.86^{+  0.14}_{-  0.14}$&ZHG $\cdot\cdot\cdot$ H92 \nl
 A750           &0.1620&$ 1893^{+ 113}_{- 113}$& & $  20.17^{+  4.29}_{-  4.29}$&C96 $\cdot\cdot\cdot$ E98 \nl
 A754           &0.0535&$ 1079^{+ 234}_{- 243}$&$ 9.00^{+ 0.50}_{- 0.50}$&$  23.07^{+  1.76}_{-  1.76}$&BMM M98 E96 \nl
 A773           &0.2170& & $ 9.29^{+ 0.69}_{- 0.60}$&$  35.10^{+  5.85}_{-  5.85}$&$\cdot\cdot\cdot$ A+F E98 \nl
 A779           &0.0230&$  473^{+  76}_{-  52}$& & $   0.28^{+  0.07}_{-  0.07}$&M97 $\cdot\cdot\cdot$ M97 \nl
 A780           &0.0552& & $ 3.57^{+ 0.10}_{- 0.10}$&$  11.80^{+  1.00}_{-  1.00}$&$\cdot\cdot\cdot$ HMS E96 \nl
 A851           &0.4510&$ 1081^{+ 194}_{- 194}$&$ 6.70^{+ 2.70}_{- 1.70}$&$  16.08^{+  0.61}_{-  0.61}$&SRE M+S S+W \nl
 A957           &0.0440&$  659^{+  88}_{-  56}$& & $   1.80^{+  0.35}_{-  0.35}$&F96 $\cdot\cdot\cdot$ E96 \nl
 A959           &0.3530& & $ 6.95^{+ 1.85}_{- 1.33}$&$  23.75^{+  3.70}_{-  3.70}$&$\cdot\cdot\cdot$ M+S B93 \nl
 A963           &0.2060&$ 1100^{+ 480}_{- 210}$&$ 6.13^{+ 0.45}_{- 0.30}$&$  22.37^{+  4.17}_{-  4.17}$&L+H A+F E98 \nl
 A1060          &0.0126&$  610^{+  52}_{-  43}$&$ 3.24^{+ 0.06}_{- 0.06}$&$   0.82^{+  0.07}_{-  0.07}$&F96 HMS E96 \nl
 A1068          &0.1390& & $ 5.50^{+ 0.90}_{- 0.90}$&$  16.04^{+  2.56}_{-  2.56}$&$\cdot\cdot\cdot$ A+F E98 \nl
 A1069          &0.0662&$  360^{+ 118}_{-  59}$& & $   2.16^{+  0.52}_{-  0.52}$&F96 $\cdot\cdot\cdot$ E96 \nl
 A1142          &0.0350&$  486^{+  81}_{-  41}$&$ 3.70^{+ 2.00}_{- 2.00}$&$   0.40^{+  0.07}_{-  0.07}$&F96 D93 J+F \nl
 A1146          &0.1422&$ 1028^{+  93}_{-  96}$& & $   6.60^{+  1.03}_{-  1.03}$&GEF $\cdot\cdot\cdot$ H92 \nl
 A1185          &0.0314&$  623^{+  65}_{-  50}$&$ 3.90^{+ 2.00}_{- 1.10}$&$   0.51^{+  0.13}_{-  0.13}$&M97 D93 E98 \nl
 A1213          &0.0468&$  549^{+ 203}_{- 105}$& &     0.59 &Q+M $\cdot\cdot\cdot$ A+K \nl
 A1240          &0.1590& & $ 3.83^{+ 0.19}_{- 0.19}$&$   2.71^{+  0.13}_{-  0.13}$&$\cdot\cdot\cdot$ M+S DFJ \nl
 A1246          &0.1870& & $ 6.26^{+ 0.54}_{- 0.49}$&$  16.70^{+  3.62}_{-  3.62}$&$\cdot\cdot\cdot$ M+S E98 \nl
 A1285          &0.1050& & $ 4.10^{+ 5.30}_{- 1.70}$&$  10.12^{+  1.92}_{-  1.92}$&$\cdot\cdot\cdot$ D93 E96 \nl
 A1291          &0.0530&   919 & & $    1.34 ^{+  0.07}_{-  0.07}$&S+R $\cdot\cdot\cdot$ J+F \nl
 A1300          &0.3071&  1200 & & $   47.63 ^{+ 12.99}_{- 13.92}$&P97 $\cdot\cdot\cdot$ E96 \nl
 A1314          &0.0329&$  664^{+ 171}_{- 105}$&$ 5.00^{+ 4.50}_{- 1.80}$&$   0.57^{+  0.13}_{-  0.13}$&DDD D93 E98 \nl
 A1367          &0.0214&$  822^{+  69}_{-  55}$&$ 3.50^{+ 0.18}_{- 0.18}$&$   2.87^{+  0.17}_{-  0.17}$&ZHG D93 E98 \nl
 A1377          &0.0514&   488 & & $    0.91 ^{+  0.05}_{-  0.05}$&S+R $\cdot\cdot\cdot$ J+F \nl
 A1413          &0.1427& & $ 8.85^{+ 0.50}_{- 0.50}$&$  36.01^{+  4.54}_{-  4.54}$&$\cdot\cdot\cdot$ A+E E98 \nl
 A1507          &0.0604&   233 & & $    0.54 ^{+  0.18}_{-  0.18}$&S+R $\cdot\cdot\cdot$ B93 \nl
 A1576          &0.3020& & $ 7.10^{+ 0.40}_{- 0.40}$&$  21.36^{+  3.19}_{-  3.19}$&$\cdot\cdot\cdot$ R99 B93 \nl
 A1631          &0.0464&$  702^{+  54}_{-  46}$& 2.80&$   0.33^{+  0.07}_{-  0.07}$&F96 $\cdot\cdot\cdot$ A+K \nl
 A1644          &0.0467&$  763^{+  64}_{-  50}$&$ 4.70^{+ 0.50}_{- 0.50}$&$   7.14^{+  0.85}_{-  0.85}$&GEF D93 E96 \nl
 A1650          &0.0840& & $ 5.50^{+ 1.30}_{- 1.00}$&$  16.83^{+  1.90}_{-  1.90}$&$\cdot\cdot\cdot$ D93 E96 \nl
 A1651          &0.0825&$  965^{+ 160}_{- 107}$&$ 6.10^{+ 0.20}_{- 0.20}$&$  18.78^{+  2.21}_{-  2.21}$&ZHG M98 E96 \nl
 A1656          &0.0231&$ 1010^{+  51}_{-  44}$&$ 8.38^{+ 0.34}_{- 0.34}$&$  20.42^{+  0.53}_{-  0.53}$&ZHG HMS E98 \nl
 A1689          &0.1810&$ 1800^{+ 200}_{- 200}$&$ 9.02^{+ 0.40}_{- 0.30}$&$  55.73^{+  8.92}_{-  8.92}$&G89 M+S E96 \nl
 A1704          &0.2190& & $ 4.44^{+ 0.73}_{- 0.52}$&$  16.09^{+  0.49}_{-  0.49}$&$\cdot\cdot\cdot$ R98 R98 \nl
 A1722          &0.3270&&$ 5.87^{+ 0.51}_{- 0.41}$&   21.00  & $\cdot\cdot\cdot$ M+S M+S \nl
 A1736          &0.0431&$  528^{+ 136}_{-  87}$&$ 4.60^{+ 0.60}_{- 0.50}$&$   4.77^{+  0.62}_{-  0.62}$&BMM D93 E96 \nl
 A1758          &0.2790& & $ 9.33^{+ 1.57}_{- 1.22}$&$  28.66^{+  1.12}_{-  1.12}$&$\cdot\cdot\cdot$ R98 R98 \nl
 A1758N         &0.2800&&$10.19^{+ 2.29}_{- 1.67}$&   43.90  & $\cdot\cdot\cdot$ M+S M+S \nl
 A1763          &0.1870& & $ 8.98^{+ 1.02}_{- 0.84}$&$  39.89^{+  5.70}_{-  5.70}$&$\cdot\cdot\cdot$ M+S E98 \nl
 A1767          &0.0700&$  933^{+ 232}_{- 134}$&$ 4.10^{+ 1.70}_{- 1.10}$&$   4.64^{+  0.54}_{-  0.54}$&ZHG D93 E98 \nl
 A1775          &0.0696&$ 1522^{+ 570}_{- 273}$&$ 4.90^{+ 2.70}_{- 1.40}$&$   5.96^{+  0.85}_{-  0.85}$&Q+M D93 E98 \nl
 A1795          &0.0631&$  828^{+  88}_{-  72}$&$ 5.88^{+ 0.14}_{- 0.14}$&$  25.42^{+  1.47}_{-  1.47}$&GEF HMS E96 \nl
 A1800          &0.0748&   724 & & $    6.91 ^{+  0.97}_{-  0.97}$&S+R $\cdot\cdot\cdot$ E98 \nl
 A1809          &0.0789&$  765^{+  79}_{-  66}$& & $   3.64^{+  0.80}_{-  0.80}$&F96 $\cdot\cdot\cdot$ E98 \nl
 A1831          &0.0612&   316 & & $    4.34 ^{+  0.57}_{-  0.57}$&S+R $\cdot\cdot\cdot$ E98 \nl
 A1835          &0.2523& & $ 9.80^{+ 1.40}_{- 1.40}$&$ 104.78^{+ 14.33}_{- 14.33}$&$\cdot\cdot\cdot$ A+F E98 \nl
 A1837          &0.0376& & $ 2.40^{+ 0.90}_{- 0.80}$&$   2.56^{+  0.62}_{-  0.62}$&$\cdot\cdot\cdot$ E+S E96 \nl
 A1904          &0.0714&$  724^{+ 149}_{-  94}$& & $   1.33^{+  0.07}_{-  0.07}$&Q+M $\cdot\cdot\cdot$ DFJ \nl
 A1913          &0.0527&$  454^{+ 128}_{-  75}$& & $   0.79^{+  0.04}_{-  0.04}$&ZHG $\cdot\cdot\cdot$ J+F \nl
 A1940          &0.1384&$  534^{+ 176}_{-  93}$& & $   3.52^{+  0.23}_{-  0.23}$&DDD $\cdot\cdot\cdot$ DFJ \nl
 A1983          &0.0452&$  551^{+  71}_{-  47}$& & $   1.08^{+  0.26}_{-  0.26}$&GEF $\cdot\cdot\cdot$ E98 \nl
 A1991          &0.0586&$  658^{+ 228}_{- 114}$&$ 5.40^{+ 5.90}_{- 2.20}$&$   2.98^{+  0.50}_{-  0.50}$&GBG D93 E98 \nl
 A1995          &0.3180&&$10.70^{+ 2.50}_{- 1.80}$&   33.30  & $\cdot\cdot\cdot$ M+S M+S \nl
 A2009          &0.1530&  804&$ 7.80^{+ 2.10}_{- 4.40}$&$  22.67^{+  4.27}_{-  4.27}$&S+R D93 E96 \nl
 A2029          &0.0765&$ 1164^{+  98}_{-  78}$&$ 8.47^{+ 0.41}_{- 0.36}$&$  41.93^{+  2.96}_{-  2.96}$&F96 A+F E98 \nl
 A2040          &0.0456&$  458^{+ 141}_{- 102}$& & $   0.88^{+  0.04}_{-  0.04}$&F96 $\cdot\cdot\cdot$ J+F \nl
 A2052          &0.0348&$  561^{+  87}_{-  73}$&$ 3.10^{+ 0.20}_{- 0.20}$&$   4.27^{+  0.34}_{-  0.34}$&GEF D93 E98 \nl
 A2055          &0.0530&&  5.80 &$  10.85^{+  1.85}_{-  1.85}$&$\cdot\cdot\cdot$ D93 E98 \nl
 A2061          &0.0777&$  554^{+ 131}_{-  77}$& & $   8.87^{+  1.29}_{-  1.29}$&ZHG $\cdot\cdot\cdot$ E98 \nl
 A2063          &0.0355&$  667^{+  55}_{-  41}$&$ 3.68^{+ 0.11}_{- 0.11}$&$   3.69^{+  0.36}_{-  0.36}$&F96 HMS E98 \nl
 A2065          &0.0722&$ 1108^{+ 273}_{- 349}$&$ 8.40^{+ 1.70}_{- 1.20}$&$  13.52^{+  1.70}_{-  1.70}$&DDD D93 E98 \nl
 A2069          &0.1160&   831 & & $   19.78 ^{+  2.79}_{-  2.79}$&S+R $\cdot\cdot\cdot$ E98 \nl
 A2079          &0.0656&$  670^{+ 113}_{-  67}$& & $   2.71^{+  0.50}_{-  0.50}$&F96 $\cdot\cdot\cdot$ A+K \nl
 A2092          &0.0670&$  504^{+ 115}_{-  69}$& & $   0.90^{+  0.16}_{-  0.16}$&ZHG $\cdot\cdot\cdot$ H92 \nl
 A2107          &0.0421&$  577^{+ 177}_{- 127}$&$ 3.78^{+ 0.19}_{- 0.19}$&$   2.02^{+  0.31}_{-  0.31}$&BMM HMS E98 \nl
 A2111          &0.2290& & $ 5.38^{+ 0.50}_{- 0.47}$&$  21.79^{+  5.28}_{-  5.28}$&$\cdot\cdot\cdot$ HWU E98 \nl
 A2124          &0.0654&$  809^{+  73}_{-  60}$& & $   3.07^{+  0.71}_{-  0.71}$&GEF $\cdot\cdot\cdot$ E98 \nl
 A2142          &0.0899&$ 1132^{+ 110}_{-  92}$&$ 9.70^{+ 1.30}_{- 1.30}$&$  61.12^{+  3.95}_{-  3.95}$&F96 HMS E96 \nl
 A2147          &0.0356&$ 1074^{+ 292}_{- 162}$&$ 4.40^{+ 0.20}_{- 0.20}$&$   5.61^{+  0.40}_{-  0.40}$&DDD D93 E96 \nl
 A2151          &0.0370&$  827^{+  69}_{-  55}$&$ 3.80^{+ 0.70}_{- 0.50}$&$   1.61^{+  0.20}_{-  0.20}$&ZHG D93 E96 \nl
 A2152          &0.0374&$  715^{+  81}_{-  61}$& &     0.58 &B+H $\cdot\cdot\cdot$ Q+M \nl
 A2162          &0.0323&$  362^{+  67}_{-  43}$& & $   0.21^{+  0.05}_{-  0.05}$&M97 $\cdot\cdot\cdot$ M97 \nl
 A2163          &0.2030& 1680&$14.60^{+ 0.85}_{- 0.85}$&$ 132.91^{+ 19.42}_{- 19.42}$&SNK A+E E96 \nl
 A2197          &0.0305&$  564^{+  84}_{-  59}$& & $   0.32^{+  0.06}_{-  0.06}$&ZHG $\cdot\cdot\cdot$ A+K \nl
 A2199          &0.0299&$  794^{+  77}_{-  60}$&$ 4.10^{+ 0.08}_{- 0.08}$&$   7.09^{+  0.25}_{-  0.25}$&ZHG HMS E98 \nl
 A2204          &0.1523& & $ 9.20^{+ 1.50}_{- 1.50}$&$  58.63^{+  6.35}_{-  6.35}$&$\cdot\cdot\cdot$ A+F E98 \nl
 A2218          &0.1710&$ 1370^{+ 160}_{- 210}$&$ 7.10^{+ 0.20}_{- 0.20}$&$  21.96^{+  2.01}_{-  2.01}$&LPS A+F E98 \nl
 A2219          &0.2280& & $12.40^{+ 0.50}_{- 0.50}$&$  64.56^{+  6.96}_{-  6.96}$&$\cdot\cdot\cdot$ A+F E98 \nl
 A2244          &0.0970& 1240&$ 8.47^{+ 0.43}_{- 0.42}$&$  25.32^{+  2.14}_{-  2.14}$&S+R M+S E98 \nl
 A2246          &0.2500&&$ 5.20^{+ 2.60}_{- 2.60}$&    1.19  & $\cdot\cdot\cdot$ R97 R97 \nl
 A2246B         &0.4400&&$ 2.60^{+ 1.30}_{- 1.30}$&    1.46  & $\cdot\cdot\cdot$ R97 R97 \nl
 A2255          &0.0808&$ 1221^{+ 181}_{- 126}$&$ 7.30^{+ 1.10}_{- 1.70}$&$  12.42^{+  0.55}_{-  0.55}$&DDD D93 E98 \nl
 A2256          &0.0581&$ 1348^{+  86}_{-  64}$&$ 7.51^{+ 0.19}_{- 0.19}$&$  18.39^{+  0.80}_{-  0.80}$&F96 D93 E98 \nl
 A2271          &0.0568&   460 & & $    1.23 ^{+  0.09}_{-  0.09}$&S+R $\cdot\cdot\cdot$ J+F \nl
 A2280          &0.3260&$  948^{+ 516}_{- 285}$& &    16.76 &G95 $\cdot\cdot\cdot$ G95 \nl
 A2319          &0.0559&$ 1545^{+  95}_{-  77}$&$ 9.12^{+ 0.15}_{- 0.15}$&$  39.74^{+  2.17}_{-  2.17}$&F96 A+E E96 \nl
 A2390          &0.2279&$ 1093^{+  61}_{-  61}$&$11.10^{+ 1.00}_{- 1.00}$&$  63.49^{+ 14.87}_{- 14.87}$&C96 B98 E98 \nl
 A2420          &0.0838& & $ 6.00^{+ 2.30}_{- 1.20}$&$  12.91^{+  2.43}_{-  2.43}$&$\cdot\cdot\cdot$ D93 D99 \nl
 A2426          &0.0886&$  332^{+  80}_{-  28}$& & $  13.43^{+  1.72}_{-  1.72}$&F96 $\cdot\cdot\cdot$ D99 \nl
 A2440          &0.0904&$  991^{+ 200}_{- 117}$& 9.00&$  11.13^{+  2.59}_{-  2.59}$&GBG D93 E96 \nl
 A2507          &0.1960& & $ 9.40^{+ 1.60}_{- 1.20}$&$  21.65^{+  0.49}_{-  0.49}$&$\cdot\cdot\cdot$ D93 R98 \nl
 A2556          &0.0865&$ 1247^{+ 249}_{- 249}$& & $   5.56^{+  2.36}_{-  2.36}$&D78 $\cdot\cdot\cdot$ E96 \nl
 A2589          &0.0416&$  500^{+ 110}_{-  67}$&$ 3.70^{+ 1.30}_{- 0.70}$&$   3.42^{+  0.38}_{-  0.38}$&ZHG D93 E98 \nl
 A2593          &0.0433&$  710^{+ 113}_{-  68}$&$ 3.10^{+ 1.50}_{- 0.90}$&$   1.98^{+  0.29}_{-  0.29}$&GBG D93 E98 \nl
 A2597          &0.0852& & $ 4.40^{+ 0.40}_{- 0.70}$&$  15.37^{+  1.79}_{-  1.79}$&$\cdot\cdot\cdot$ M98 E96 \nl
 A2626          &0.0573&$  658^{+ 111}_{-  81}$&$ 2.90^{+ 2.50}_{- 1.00}$&$   3.21^{+  0.44}_{-  0.44}$&M+W D93 E98 \nl
 A2634          &0.0309&$  700^{+  97}_{-  61}$&$ 3.07^{+ 0.11}_{- 0.11}$&$   1.59^{+  0.17}_{-  0.17}$&F96 A+E E96 \nl
 A2657          &0.0414&  667&$ 3.40^{+ 0.50}_{- 0.30}$&$   2.82^{+  0.35}_{-  0.35}$&S+R D93 E98 \nl
 A2666          &0.0270&$  476^{+  95}_{-  60}$& &     0.05 &ZHG $\cdot\cdot\cdot$ Q+M \nl
 A2670          &0.0759&$  918^{+  65}_{-  47}$&$ 4.45^{+ 0.20}_{- 0.20}$&$   4.97^{+  0.92}_{-  0.92}$&GEF HMS E96 \nl
 A2717          &0.0498&$  541^{+  65}_{-  41}$& & $   2.32^{+  0.39}_{-  0.39}$&F96 $\cdot\cdot\cdot$ E96 \nl
 A2721          &0.1152&$  805^{+  74}_{-  63}$& & $   8.48^{+  1.70}_{-  1.70}$&F96 $\cdot\cdot\cdot$ E96 \nl
 A2734          &0.0625&$  628^{+  61}_{-  57}$& & $   5.82^{+  0.87}_{-  0.87}$&F96 $\cdot\cdot\cdot$ E96 \nl
 A2744          &0.3080&$ 1950^{+ 334}_{- 334}$&$11.00^{+ 0.50}_{- 0.50}$&$  62.44^{+ 14.41}_{- 14.41}$&FMB A+F E96 \nl
 A2877          &0.0248&$  744^{+  63}_{-  51}$&$ 3.50^{+ 1.10}_{- 0.80}$&$   0.82^{+  0.13}_{-  0.13}$&GEF D93 E96 \nl
 A3093          &0.0836&$  440^{+  80}_{-  56}$& & $   0.21^{+  0.02}_{-  0.02}$&F96 $\cdot\cdot\cdot$ DFJ \nl
 A3112          &0.0746&$  552^{+  86}_{-  63}$&$ 4.24^{+ 0.24}_{- 0.24}$&$  14.67^{+  1.20}_{-  1.20}$&F96 D99 E96 \nl
 A3126          &0.0862&$ 1053^{+ 164}_{- 108}$& & $   7.18^{+  1.08}_{-  1.08}$&F96 $\cdot\cdot\cdot$ E96 \nl
 A3128          &0.0604&$  841^{+  51}_{-  44}$& & $   4.84^{+  0.50}_{-  0.50}$&GEF $\cdot\cdot\cdot$ E96 \nl
 A3158          &0.0575&$  976^{+  70}_{-  58}$&$ 5.50^{+ 0.30}_{- 0.40}$&$  11.60^{+  0.74}_{-  0.74}$&F96 D93 E96 \nl
 A3223          &0.0603&$  647^{+  67}_{-  54}$& & $   1.87^{+  0.05}_{-  0.05}$&F96 $\cdot\cdot\cdot$ DFJ \nl
 A3266          &0.0594&$ 1138^{+  94}_{-  74}$&$ 8.00^{+ 0.30}_{- 0.30}$&$  16.48^{+  0.64}_{-  0.64}$&GEF M98 E96 \nl
 A3360          &0.0849&$  835^{+ 114}_{-  82}$& & $   2.07^{+  0.17}_{-  0.17}$&F96 $\cdot\cdot\cdot$ DFJ \nl
 A3376          &0.0490&$  723^{+  82}_{-  61}$&$ 4.00^{+ 0.40}_{- 0.40}$&$   4.66^{+  0.41}_{-  0.41}$&GEF HMS E96 \nl
 A3389          &0.0267&$  595^{+  63}_{-  47}$&$ 2.10^{+ 0.90}_{- 0.60}$&$   0.48^{+  0.03}_{-  0.03}$&F96 D93 E96 \nl
 A3391          &0.0553&$  786^{+  78}_{-  53}$&$ 5.40^{+ 0.60}_{- 0.60}$&$   5.03^{+  0.41}_{-  0.41}$&GEF HMS E96 \nl
 A3395          &0.0506&$  823^{+  51}_{-  43}$&$ 5.00^{+ 0.30}_{- 0.30}$&$   5.85^{+  0.40}_{-  0.40}$&GEF HMS E96 \nl
 A3526          &0.0114&$  586^{+  45}_{-  35}$&$ 3.68^{+ 0.06}_{- 0.06}$&$   1.53^{+  0.07}_{-  0.07}$&ZHG HMS E96 \nl
 A3528N         &0.0553&$  972^{+ 110}_{-  82}$& &     0.49 &F96 $\cdot\cdot\cdot$ Q+M \nl
 A3532          &0.0559&$  738^{+ 112}_{-  85}$&$ 4.40^{+ 4.70}_{- 1.30}$&$   5.54^{+  0.82}_{-  0.82}$&F96 E90 E96 \nl
 A3556          &0.0476&$  580^{+ 100}_{-  73}$& & $   2.16^{+  0.53}_{-  0.53}$&F96 $\cdot\cdot\cdot$ E96 \nl
 A3558          &0.0475&$  735^{+  49}_{-  41}$&$ 5.12^{+ 0.20}_{- 0.20}$&$  13.27^{+  1.16}_{-  1.16}$&GEF HMS E96 \nl
 A3559          &0.0469&$  456^{+  78}_{-  44}$& & $   0.45^{+  0.03}_{-  0.03}$&F96 $\cdot\cdot\cdot$ DFJ \nl
 A3562          &0.0478&$  736^{+  49}_{-  36}$&$ 3.80^{+ 0.50}_{- 0.50}$&$   6.11^{+  0.77}_{-  0.77}$&F96 D93 E96 \nl
 A3571          &0.0396&$ 1045^{+ 109}_{-  90}$&$ 6.73^{+ 0.17}_{- 0.17}$&$  18.09^{+  1.20}_{-  1.20}$&F96 HMS E96 \nl
 A3651          &0.0610&$  626^{+  60}_{-  53}$& & $   3.35^{+  1.41}_{-  1.41}$&F96 $\cdot\cdot\cdot$ D99 \nl
 A3667          &0.0566&$  971^{+  62}_{-  47}$&$ 7.00^{+ 0.60}_{- 0.60}$&$  22.70^{+  4.20}_{-  4.20}$&F96 HMS E98 \nl
 A3693          &0.0921&$  478^{+ 107}_{-  50}$& & $   4.54^{+  1.65}_{-  1.65}$&F96 $\cdot\cdot\cdot$ D99 \nl
 A3695          &0.0903&$  779^{+  67}_{-  49}$& & $  11.39^{+  1.96}_{-  1.96}$&F96 $\cdot\cdot\cdot$ E96 \nl
 A3716          &0.0493&$  954^{+ 141}_{-  85}$& & $   2.32^{+  0.41}_{-  0.41}$&GBG $\cdot\cdot\cdot$ E96 \nl
 A3733          &0.0380&$  608^{+ 109}_{-  60}$& & $   0.97^{+  0.23}_{-  0.23}$&F96 $\cdot\cdot\cdot$ E96 \nl
 A3744          &0.0387&$  508^{+  74}_{-  48}$& & $   8.85^{+  1.54}_{-  1.54}$&F96 $\cdot\cdot\cdot$ D99 \nl
 A3809          &0.0631&$  478^{+  62}_{-  45}$& & $   5.13^{+  0.89}_{-  0.89}$&F96 $\cdot\cdot\cdot$ E96 \nl
 A3822          &0.0769&$  810^{+  89}_{-  58}$& & $   9.03^{+  1.24}_{-  1.24}$&F96 $\cdot\cdot\cdot$ E96 \nl
 A3825          &0.0760&$  699^{+  79}_{-  58}$& & $   4.03^{+  0.86}_{-  0.86}$&F96 $\cdot\cdot\cdot$ E96 \nl
 A3827          &0.0984&$  962^{+ 407}_{- 407}$& & $  22.82^{+  2.52}_{-  2.52}$&AMB $\cdot\cdot\cdot$ D99 \nl
 A3880          &0.0380&$  855^{+ 148}_{- 148}$& 3.80&$   3.43^{+  0.52}_{-  0.52}$&S97 $\cdot\cdot\cdot$ E96 \nl
 A3888          &0.1680&$ 1307^{+ 100}_{-  92}$& & $  31.40^{+  4.91}_{-  4.91}$&GEF $\cdot\cdot\cdot$ E96 \nl
 A3921          &0.0944&$  490^{+ 126}_{-  73}$&$ 4.90^{+ 0.55}_{- 0.55}$&$  10.92^{+  1.52}_{-  1.52}$&F96 A+E E96 \nl
 A4010          &0.0966&$  625^{+ 127}_{-  95}$& & $  12.43^{+  4.05}_{-  4.05}$&F96 $\cdot\cdot\cdot$ E96 \nl
 A4038          &0.0302&$  898^{+ 112}_{- 116}$&$ 3.30^{+ 1.60}_{- 0.80}$&$   3.31^{+  0.26}_{-  0.26}$&GEF D93 E96 \nl
 A4059          &0.0478&$  845^{+ 280}_{- 140}$&$ 3.97^{+ 0.12}_{- 0.12}$&$   5.78^{+  0.54}_{-  0.54}$&GGP HMS E96 \nl
 1ES0657$-$558    &0.2994& & $11.00^{+ 1.50}_{- 1.30}$&$ 126.81^{+ 14.63}_{- 14.63}$&$\cdot\cdot\cdot$ Y98 D99 \nl
 2A0335+096     &0.0350& & $ 3.01^{+ 0.07}_{- 0.07}$&$   7.05^{+  0.13}_{-  0.13}$&$\cdot\cdot\cdot$ HMS E98 \nl
 3C129          &0.0218& & $ 5.60^{+ 0.40}_{- 0.40}$&$   4.31^{+  0.39}_{-  0.39}$&$\cdot\cdot\cdot$ D93 E+S \nl
 3C295          &0.4600&$ 1670^{+ 364}_{- 364}$&$ 7.13^{+ 2.06}_{- 1.35}$&$  26.00^{+  4.00}_{-  2.00}$&SRE M+S N99 \nl
 AC114          &0.3100&$ 1649^{+ 217}_{- 156}$&$ 9.76^{+ 1.04}_{- 0.85}$&  38.10&FMB A+F A+F \nl
 AWM4           &0.0424& &$ 2.38^{+ 0.17}_{- 0.17}$&$   0.68^{+  0.10}_{-  0.10}$&$\cdot\cdot\cdot$ HMS E98 \nl
 AWM7           &0.0176&$  864^{+ 113}_{-  81}$&$ 3.75^{+ 0.09}_{- 0.09}$&$   3.00^{+  0.45}_{-  0.45}$&GBG HMS E+S \nl
 AXJ2019$-$1127   &0.9400& & $ 8.60^{+ 4.20}_{- 3.00}$&$  19.42^{+  5.55}_{-  3.93}$&$\cdot\cdot\cdot$ H97 H97 \nl
 CA0340$-$538     &0.0570&$ 1006^{+ 222}_{- 135}$& & $   9.16^{+  1.04}_{-  1.04}$&DDD $\cdot\cdot\cdot$ P92 \nl
 CL0016+16      &0.5545&$ 1234^{+ 128}_{- 128}$&$ 8.00^{+ 1.00}_{- 1.00}$&$  28.13^{+  3.14}_{-  3.14}$&C96 M+S H92 \nl
 CL0024+16      &0.3910&$ 1339^{+ 233}_{- 233}$& & $   4.26^{+  0.91}_{-  0.91}$&SRE $\cdot\cdot\cdot$ FMB \nl
 CL0107$-$46      &0.0230&$ 1032^{+ 125}_{- 108}$& & $   0.85^{+  0.02}_{-  0.02}$&B95 $\cdot\cdot\cdot$ J+F \nl
 CL0336+09      &0.0349& & $ 3.00^{+ 0.20}_{- 0.10}$&$   7.12^{+  0.64}_{-  0.64}$&$\cdot\cdot\cdot$ D93 E+S \nl
 CL0422$-$09      &0.0390& & $ 2.90^{+ 0.50}_{- 0.40}$&$   3.15^{+  0.73}_{-  0.73}$&$\cdot\cdot\cdot$ D93 E+S \nl
 CL0500$-$24      &0.3270&$ 1300^{+ 300}_{- 300}$&$ 7.20^{+ 3.70}_{- 1.80}$&$  17.51^{+  1.33}_{-  0.89}$&W89 OMF S+W \nl
 CL0745$-$19      &0.1028& & $ 8.50^{+ 1.90}_{- 1.40}$&$  59.00^{+ 19.10}_{- 19.10}$&$\cdot\cdot\cdot$ E+S E+S \nl
 CL1322+30      &0.7570&$  820^{+ 120}_{- 120}$& & $   1.43^{+  0.44}_{-  0.44}$&C94 $\cdot\cdot\cdot$ C94 \nl
 CL1447+26      &0.3762&  1470 & & $   10.69 ^{+  1.98}_{-  1.98}$&D99 $\cdot\cdot\cdot$ FMB \nl
 CL2244$-$02      &0.3280&&$ 6.50^{+ 1.80}_{- 1.30}$&    2.91  & $\cdot\cdot\cdot$ OMF OMF \nl
 Cygnus$-$A       &0.0570&$ 1581^{+ 286}_{- 197}$&$ 6.50^{+ 0.36}_{- 0.36}$&  16.50&O97 MME D93 \nl
 IRAS09104+4109 &0.4420&&$ 8.50^{+ 3.40}_{- 3.40}$&   55.10  & $\cdot\cdot\cdot$ A+F A+F \nl
 MKW3S          &0.0434&$  603^{+  61}_{-  59}$&$ 3.00^{+ 0.30}_{- 0.30}$&$   4.47^{+  0.50}_{-  0.50}$&GEF D93 E98 \nl
 MKW4           &0.0196&$  535^{+  65}_{-  59}$&$ 1.71^{+ 0.09}_{- 0.09}$&$   0.49^{+  0.06}_{-  0.06}$&GBG HMS B96 \nl
 MKW4S          &0.0288& & $ 1.95^{+ 0.17}_{- 0.17}$&$   0.31^{+  0.06}_{-  0.06}$&$\cdot\cdot\cdot$ HMS E98 \nl
 MKW7           &0.0289&$  573^{+ 363}_{- 172}$& & $   0.15^{+  0.04}_{-  0.04}$&L96 $\cdot\cdot\cdot$ B96 \nl
 MKW8           &0.0272&$  422^{+  99}_{-  53}$& & $   1.40^{+  0.20}_{-  0.20}$&L96 $\cdot\cdot\cdot$ E98 \nl
 MKW9           &0.0397&  336&$ 2.23^{+ 0.13}_{- 0.13}$&   0.10&B99 HMS B99 \nl
 MKW10          &0.0206&$  177^{+  85}_{-  46}$& & $   0.14^{+  0.03}_{-  0.03}$&L96 $\cdot\cdot\cdot$ B96 \nl
 MKW11          &0.0232&$  384^{+  70}_{-  42}$& & $   0.29^{+  0.06}_{-  0.06}$&L96 $\cdot\cdot\cdot$ B96 \nl
 MS0302+16      &0.4246&$  646^{+  93}_{-  93}$& & $   9.08^{+  1.00}_{-  1.00}$&C96 $\cdot\cdot\cdot$ H92 \nl
 MS0302+17      &0.4250& & $ 4.60^{+ 0.80}_{- 0.80}$&$   4.23^{+  0.69}_{-  0.69}$&$\cdot\cdot\cdot$ K99 H92 \nl
 MS0353$-$36      &0.3200& & $ 8.13^{+ 2.57}_{- 1.73}$&$  14.52^{+  0.65}_{-  0.65}$&$\cdot\cdot\cdot$ M+S H97 \nl
 MS0440+02      &0.1965&$  606^{+  62}_{-  62}$&$ 5.30^{+ 1.27}_{- 0.85}$&$   7.43^{+  0.95}_{-  0.95}$&C96 M+S H92 \nl
 MS0451+02      &0.2011&  979&$ 8.60^{+ 0.50}_{- 0.50}$&$  15.93^{+  2.70}_{-  2.70}$&C96 M+S H92 \nl
 MS0451$-$03      &0.5392&$ 1371^{+ 105}_{- 105}$&$10.17^{+ 1.55}_{- 1.26}$&  53.70&C96 M+S M+S \nl
 MS0811+63      &0.3120& & $ 4.60^{+ 0.90}_{- 0.60}$&$   2.91^{+  0.58}_{-  0.58}$&$\cdot\cdot\cdot$ H97 H92 \nl
 MS0839+29      &0.1928&$  749^{+ 104}_{- 104}$&$ 4.19^{+ 0.36}_{- 0.33}$&$   9.14^{+  1.25}_{-  1.25}$&C96 M+S H92 \nl
 MS1006+12      &0.2605&$  906^{+ 101}_{- 101}$& & $   9.16^{+  1.47}_{-  1.47}$&C96 $\cdot\cdot\cdot$ H92 \nl
 MS1008$-$12      &0.3062&$ 1054^{+ 107}_{- 107}$&$ 7.29^{+ 2.45}_{- 1.52}$&$   9.13^{+  1.24}_{-  1.24}$&C96 M+S H92 \nl
 MS1054$-$03      &0.8260&$ 1170^{+ 150}_{- 150}$&$12.30^{+ 3.10}_{- 2.20}$&  19.91&T99 D98 G+L \nl
 MS1147+11      &0.3030& & $ 5.50^{+ 0.80}_{- 0.60}$&$   4.19^{+  1.00}_{-  1.00}$&$\cdot\cdot\cdot$ H97 H92 \nl
 MS1224+20      &0.3255&$  802^{+  90}_{-  90}$&$ 4.30^{+ 0.70}_{- 0.60}$&$   8.06^{+  0.52}_{-  0.52}$&C96 H97 H97 \nl
 MS1231+15      &0.2350&$  667^{+  69}_{-  69}$& & $   5.52^{+  1.29}_{-  1.29}$&C96 $\cdot\cdot\cdot$ H92 \nl
 MS1241+17      &0.3120& & $ 6.20^{+ 1.80}_{- 1.30}$&$   6.47^{+  1.65}_{-  1.65}$&$\cdot\cdot\cdot$ H97 H92 \nl
 MS1305+29      &0.2410& & $ 2.98^{+ 0.57}_{- 0.41}$&$   2.24^{+  0.33}_{-  0.33}$&$\cdot\cdot\cdot$ M+S H92 \nl
 MS1358+62      &0.3283&$  937^{+  54}_{-  54}$&$ 7.50^{+ 4.30}_{- 4.30}$&$  21.81^{+  3.81}_{-  3.81}$&C96 A+F H92 \nl
 MS1426+01      &0.3200& & $ 5.50^{+ 1.10}_{- 0.70}$&$   6.71^{+  0.85}_{-  0.85}$&$\cdot\cdot\cdot$ H97 H92 \nl
 MS1455+22      &0.2570&$ 1133^{+ 140}_{- 140}$&$ 5.45^{+ 0.29}_{- 0.28}$&$  29.42^{+  1.45}_{-  1.45}$&C96 M+S H92 \nl
 MS1512+36      &0.3726&$  690^{+  96}_{-  96}$&$ 3.57^{+ 1.33}_{- 0.74}$&$   7.62^{+  1.67}_{-  1.67}$&C96 M+S H92 \nl
 MS1621+26      &0.4274&$  793^{+  55}_{-  55}$& & $   8.22^{+  1.55}_{-  1.55}$&C96 $\cdot\cdot\cdot$ H92 \nl
 MS2137$-$23      &0.3130&  960&$ 4.37^{+ 0.38}_{- 0.72}$&$  26.27^{+  3.59}_{-  3.59}$&K95 M+S H92 \nl
 Ophiuchus      &0.0280& & $ 9.10^{+ 0.30}_{- 0.30}$&$  32.20^{+  5.20}_{-  5.20}$&$\cdot\cdot\cdot$ A+E E+S \nl
 PKS0745$-$191    &0.1028&&$ 8.70^{+ 1.60}_{- 1.20}$&   64.05  & $\cdot\cdot\cdot$ A+F A98 \nl
 RXJ0658$-$5557   &0.3100& & $17.00^{+ 4.00}_{- 4.00}$&$ 126.60^{+  5.52}_{-  5.52}$&$\cdot\cdot\cdot$ A99 A99 \nl
 RXJ1347$-$1145   &0.4510& & $11.37^{+ 1.10}_{- 0.92}$&$ 197.71^{+ 21.67}_{- 21.67}$&$\cdot\cdot\cdot$ M+S SHN \nl
 RXJ1716.6+6708 &0.8130&$ 1522^{+ 215}_{- 150}$&$ 5.66^{+ 1.37}_{- 0.58}$&$  17.40^{+  0.91}_{-  0.91}$&G99 G99 G99 \nl
 S84            &0.1086&$  329^{+  60}_{-  25}$& & $   6.03^{+  1.16}_{-  1.16}$&F96 $\cdot\cdot\cdot$ D99 \nl
 S301           &0.0223&$  506^{+ 223}_{- 125}$& & $   0.38^{+  0.08}_{-  0.08}$&GBG $\cdot\cdot\cdot$ D99 \nl
 S805           &0.0141&$  470^{+  66}_{- 103}$&$ 1.40^{+ 0.30}_{- 0.30}$&   0.20&GBG D93 D93 \nl
 S987           &0.0717&$  677^{+ 141}_{-  66}$& & $   2.85^{+  1.41}_{-  1.41}$&F96 $\cdot\cdot\cdot$ D99 \nl
 S1101          &0.0580& & $ 3.00^{+ 1.20}_{- 0.70}$&$   5.35^{+  0.50}_{-  0.50}$&$\cdot\cdot\cdot$ E+S D99 \nl
 SC0316$-$444     &0.0730&$  788^{+ 174}_{- 108}$& &    13.08 &Q+M $\cdot\cdot\cdot$ Q+M \nl
 SC1327-312     &0.0495&$  580^{+ 119}_{- 118}$&$ 3.85^{+ 2.40}_{- 1.32}$&   3.14&D97 BZM K+B \nl
 SC2008$-$569     &0.0530&$ 1470^{+ 291}_{- 183}$& & $  27.84^{+  1.83}_{-  1.83}$&Q+M $\cdot\cdot\cdot$ Q+M \nl
 SC2059$-$25      &0.1880&&$ 7.00^{+ 4.20}_{- 1.30}$&   25.30  & $\cdot\cdot\cdot$ D93 D93 \nl
 SC2311$-$43      &0.0556&&$ 2.50^{+ 0.90}_{- 0.60}$&    4.71  & $\cdot\cdot\cdot$ D93 D93 \nl
 Triangulum-Aust&0.0510& & $10.05^{+ 0.69}_{- 0.69}$&$  26.90^{+  4.80}_{-  4.80}$&$\cdot\cdot\cdot$ HMS E+S \nl
 Virgo          &0.0038&$  673^{+  48}_{-  40}$&$ 2.20^{+ 0.02}_{- 0.02}$&$   1.52^{+  0.03}_{-  0.03}$&DDD A+E E98 \nl
 WP23           &0.0087&&$ 1.00^{+ 0.60}_{- 0.40}$&    0.23  & $\cdot\cdot\cdot$ D93 D93 \nl
 ZW0628+25      &0.0810&&$ 6.20^{+ 3.60}_{- 1.70}$&    7.86  & $\cdot\cdot\cdot$ D93 D93 \nl
 ZW1615+35      &0.0321&  584&$ 2.90^{+ 2.60}_{- 1.10}$&   0.40&U78 D93 D93 \nl
 ZW3146         &0.2906& & $ 6.35^{+ 0.37}_{- 0.34}$&$  55.81^{+  9.49}_{-  9.49}$&$\cdot\cdot\cdot$ M+S E98 \nl
\tablerefs{
(respectively velocity dispersion, temperature and 
X-ray luminosity): 
A98 --- Allen (1998);
A99 --- Andreani  et al. (1999);
A+E --- Arnaud \& Evrard (1999);
A+F --- Allen \& Fabian (1998); 
A+K --- Abramopoulos  \& Ku (1983);
AMB --- Adami et al. (1998);
B93 --- Briel \& Henry (1993);
B95 --- Bird (1995);
B96 --- Burns et al. (1996);
B98 --- B\"ohringer et al. (1998);
B99 --- Buote (1999);
B+H --- Barmby \& Huchra  (1998);
BMM --- Bird, Mushotzky \& Metzler (1995);
BZM --- Bardelli et al. (1996);
C94 --- Castander et al. (1994);
C96 --- Carlberg et al. (1996);
D78 --- Dressler (1978);
D93 --- David et al. (1993);
DJF --- David, Jones \& Forman (1995);
D98 --- Donahue et al. (1998);
D99 --- De Grandi et al. (1999);
DDD --- Danese, De Zotti \& di Tullio (1980); 
DFJ --- David, Forman \& Jones (1999);
DSP --- Dressler et al. (1999);
E90 --- Edge et al. (1990);
E96 --- Ebeling et al. (1996);
E98 --- Ebeling et al. (1998);
E+S --- Edge \& Stewart (1991a);
F96 --- Fadda et al. (1996);
FMB --- Fabricant, McClintock \& Bautz  (1991);
G95 --- Gioia et al. (1995);
G89 --- Gudehus (1989);
G97 --- G\'omez et al. (1997);
G99 --- Gioia et al. (1999);
GBG --- Girardi et al. (1993);
GEF --- Girardi et al. (1997);
GGP --- Green, Godwin  \& Peach (1988)
G+L --- Gioia \& Luppino (1994);
H92 --- Henry et al. (1992);
H97 --- Henry (1997);
HMS --- Horner, Mushotzky \& Scharf (1999);
HWU --- Henriksen, Wang \& Ulmer (1999);
J+F --- Jones \& Forman (1984);
K95 --- Kneib (1995);
K99 --- Kaiser et al. (1999);
K+B --- Kull \& B\"ohringer (1998);
L96 --- Ledlow et al. (1996);
L+H --- Lavery \& Henry (1988);
LPS --- Le Borgne, Pell\'o  \& Sanahuja (1992);
M97 --- Mahdavi et al. (1997);
M98 --- Markevitch et al. (1998);
MME --- Mohr, Mathiesen \& Evrard (1999);
M+S --- Mushotzky \& Scharf (1997);
M+W --- Mohr \& Wegner (1997);
N99 --- Neumann (1999);
O97 --- Owen et al. (1997);
OMF --- Ota, Mitsuda \& Fukazawa (1998);
P92 --- Piccinotti et al. (1992);
P97 --- Pierre et al. (1997);
Q+M --- Quintana  \& Melnick (1982);
R97 --- Reimers et al. (1997);
R98 --- Rizza et al. (1998);
R99 --- Rines et al. (1999);
S97 --- Stein (1996);
S99 --- Schindler et al. (1999);
SHN --- Schindler et al. (1997);
SNK --- Squires et al. (1997);
SRE --- Smail et al. (1997):
S+R --- Struble \& Rood (1991);
S+W --- Schindler \& Wambsganss (1997);
T99 --- Tran et al. (1999);
U78 --- Ulrich (1978);
W89 --- Wambsganss et al. (1989);
WQI --- Way, Quintana \&Infante (1997);
Y98 --- Yaqoon (1998);
ZHG --- Zabludoff, Huchra \& Geller (1990).
}

 \enddata
 \end{deluxetable}


\setcounter{page}{31}

\begin{center}

{\normalsize Table 2~ Summary of the best fitted $L$-$T$ relationships}

\begin{tabular}{lcll}
 & & &  \\
\hline\hline
authors & cluster No. &  method & fitted relation  \\
\hline
Henry \& Arnaud (1991) & 24 & OLS &
		$L_x=10^{-0.99\pm0.29}T^{2.7\pm0.4}$\\
          &     &    &  $T=(1.05^{+0.25}_{-0.20})\;L_x^{0.265\pm0.035}$\\
Edge \& Stewart (1991a) & 45 & OLS & 
		$L_x=10^{-0.95\pm0.08}T^{2.62\pm0.10}$\\
          &     &    &  $T=(2.95^{+2.06}_{-1.21})\;L_x^{0.30\pm0.05}$\\
David et al (1993) & 104 & OLS &
			$T=(2.94^{+0.28}_{-0.26})\;L_x^{0.297\pm0.004}$ \\
White et al (1997) & 86 & ORD & 
		$L_x=(4.78\pm0.99)\times10^{-2}T^{2.98\pm0.11}$\\
          &     &    &  $T=(2.76\pm0.08)L_x^{0.33\pm0.01}$\\
Allen \& Fabian (1998) & 30 & BCES$^a$ & 
		$T=(1.66\pm0.52)L_x^{0.429\pm0.079}$ \\
Arnaud \& Evrard (1998) & 24 & OLS & 
		$L_x=10^{1.06\pm0.03}(T/6)^{2.88\pm0.15}$ \\
Markevitvh (1998)$^b$ & 30 & BCES & 
		$L_x=(12.44\pm1.08)(T/6)^{2.64\pm0.27}$ \\
Jone \& Forman (1999)$^c$ & 78  & OLS &
		$T=(0.13^{+0.08}_{-0.07})L_x^{0.531\pm0.068}$ \\
This work & 168 & OLS & 
	        $L_x=10^{-0.92\pm0.09}T^{2.61\pm0.12}$  \\ 
          &     &    &  $T=10^{0.45\pm0.01}L_x^{0.28\pm0.01}$\\
          & 142 & ORD & 
      		$L_x=10^{-0.92\pm0.05}T^{2.72\pm0.05}$  \\ 
          &     &    &  $T=10^{0.34\pm0.01}L_x^{0.37\pm0.01}$\\
\hline
\end{tabular}
\end{center}
\footnotesize
\noindent\null\qquad $^{a}${The modified OLS method - 
			   bivariate correlated errors and 
			   intrinsic scatter [see Akritas \& 
			   Bershady (1996)];}\\
\null\qquad $^{b}${Cooling flow regions are excluded;}\\
\null\qquad $^{c}${$L_x$, in units of $10^{40}$ erg/s, is the 0.5--4.5 keV 
		  luminosity and measured within 1 Mpc.}
\\

\clearpage

\figcaption{The $L_x$-$T$ relationship for 168 clusters whose $L_x$ and
$T$ are observationally determined (Table 1). 
The low-redshift ($z<0.1$) and high-redshift
($z\geq0.1$) clusters are represented by the open squares (100) and the filled
circles (68), respectively. The dashed line is the best OLS fitted 
relationship to the whole data set, $L_x=0.12 T^{2.61}$. 
\label{fig1}}

\figcaption{The $L_x$-$T$ relationship for a subsample of
142 clusters for which the
measurement uncertainties in both $L_x$ and $T$ are known. The symbols have
the same meaning as in Fig.1. The solid and dashed lines represent the best
fitted results by the ODR and OLS methods, respectively.
\label{fig2}}

\figcaption{The $L_x$-$\sigma$ relationship for 193 clusters 
whose $L_x$ and $\sigma$ are given in Table 1. 
The low-redshift ($z<0.1$) and high-redshift
($z\geq0.1$) clusters are represented by the open squares (147) and the filled
circles (46), respectively. The dashed line is the best OLS fitted 
relationship to the whole data set, $L_x=10^{-7.06} \sigma^{2.67}$. 
\label{fig3}}

\figcaption{The $L_x$-$\sigma$ relationship for a subsample of 156 clusters
for which the measurement uncertainties in both $L_x$ and $\sigma$ are known.
The dashed line shows the best OLS fit to the data, and the solid line
is the doubly weighted ODR result.
\label{fig4}}

\figcaption{
The $\sigma$-$T$ relationship for the 105 clusters in Table 1. The symbols
have the same meaning as in Fig.1. The dashed line shows the OLS fitted 
relationship to the data set and the solid one is the ODR result for
92 clusters that have measurement uncertainties in both $\sigma$ and $T$. 
\label{fig5}}

\figcaption{Mean cluster X-ray luminosity $L_x$, temperature $T$ and
velocity dispersion $\sigma$ are plotted against $(1+z)$ for our
cluster sample. Each redshift bin contains 16, 14 and 19
clusters for ($L_x,z$), ($T,z$) and ($\sigma,z$), respectively. 
Vertical error bar demonstrates the difference between the 
maximum and minimum values observed within each bin.  
\label{fig6}}

\figcaption{The ratios of $L_x$ to (a)  $T^{3.06}$,  
(b) $T^{2.5}$ and (c) $T^{2}(1+z)^{3/2}$ 
for our cluster sample. The horizonal axis is plotted in order of 
clusters in Table 1 where their $L_x$ and $T$ are available.  
The open squares and the filled triangles
represent the low- and high-redshift clusters, respectively. The dotted lines
show the mean values. 
\label{fig7}}

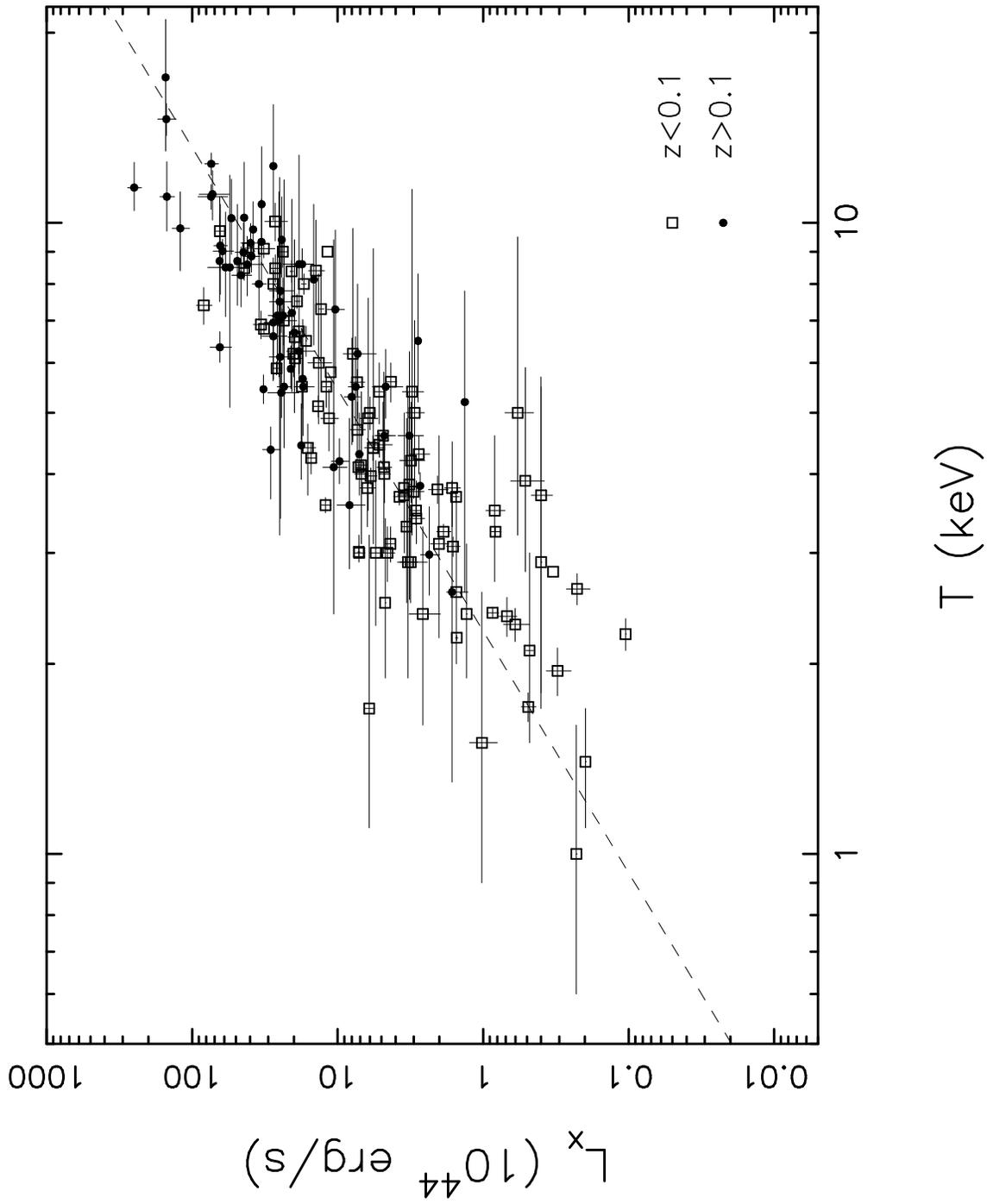
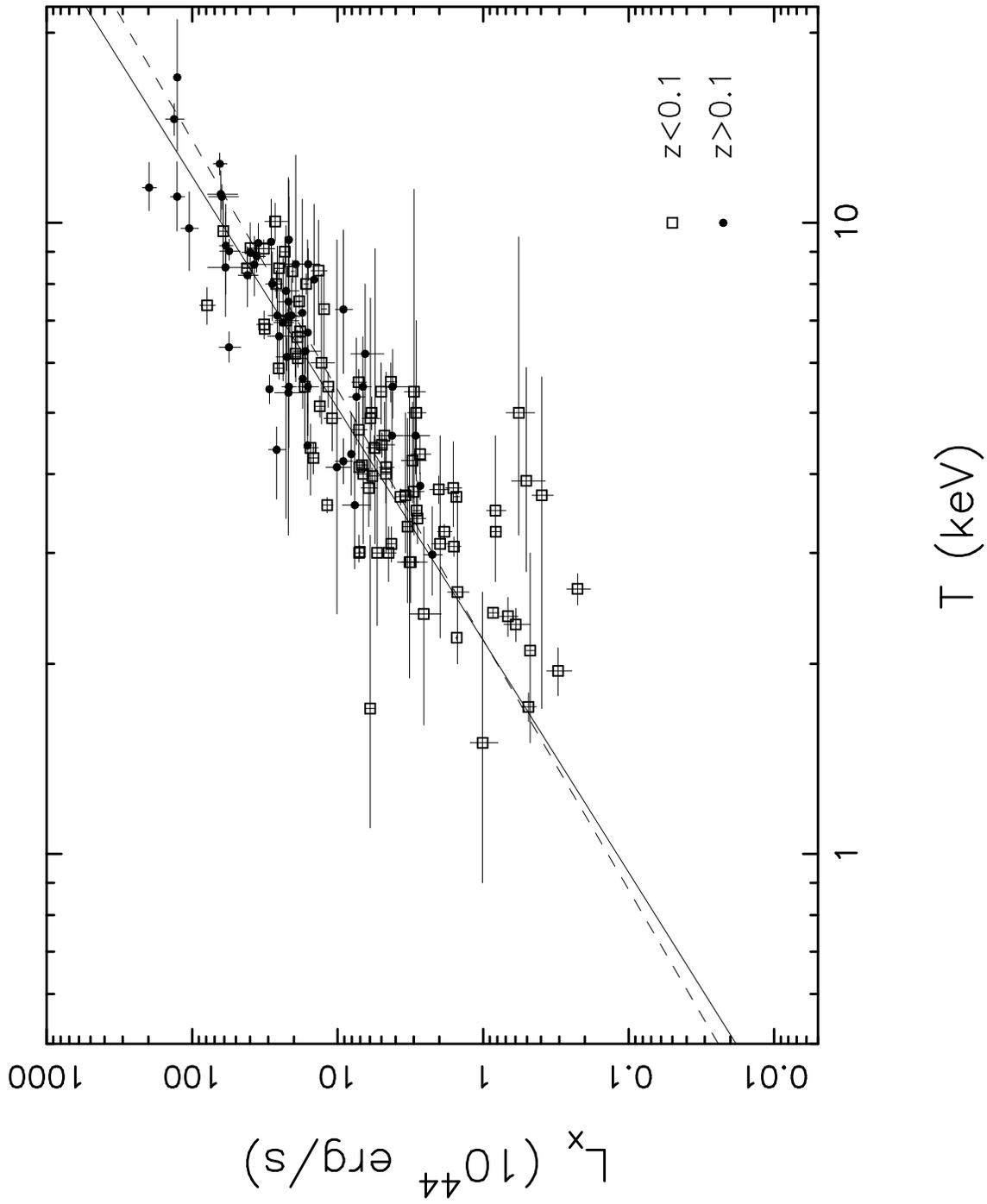
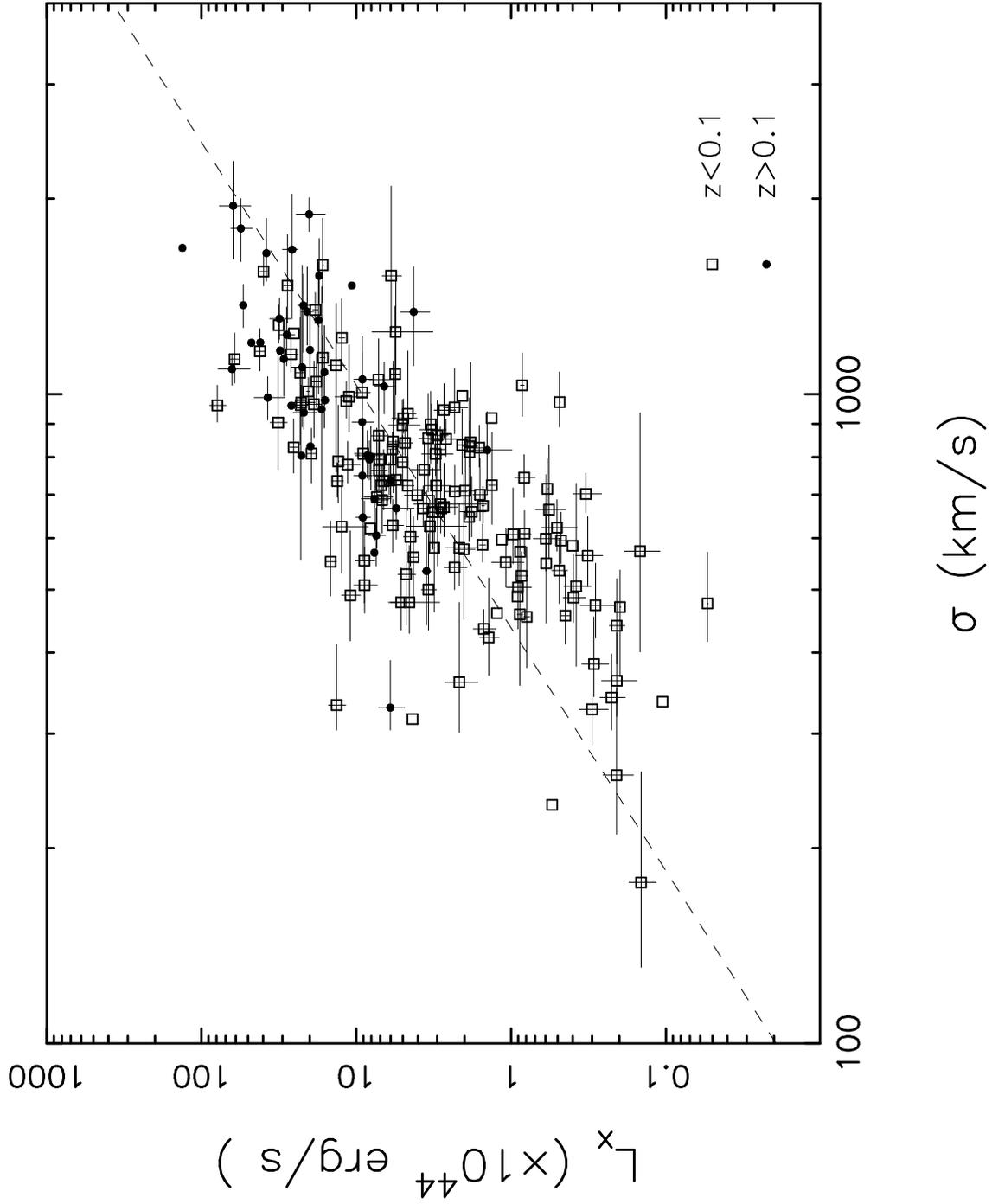
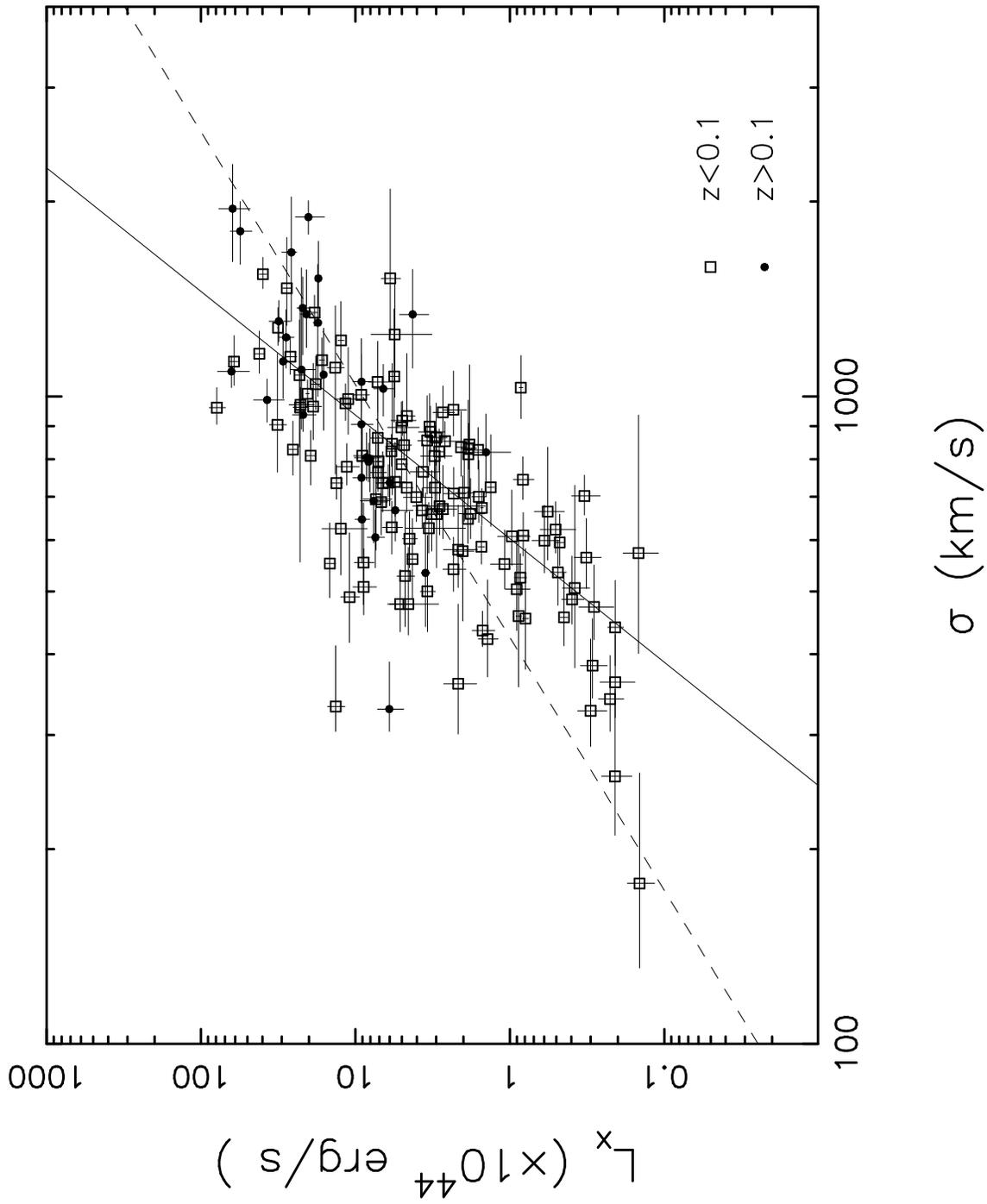
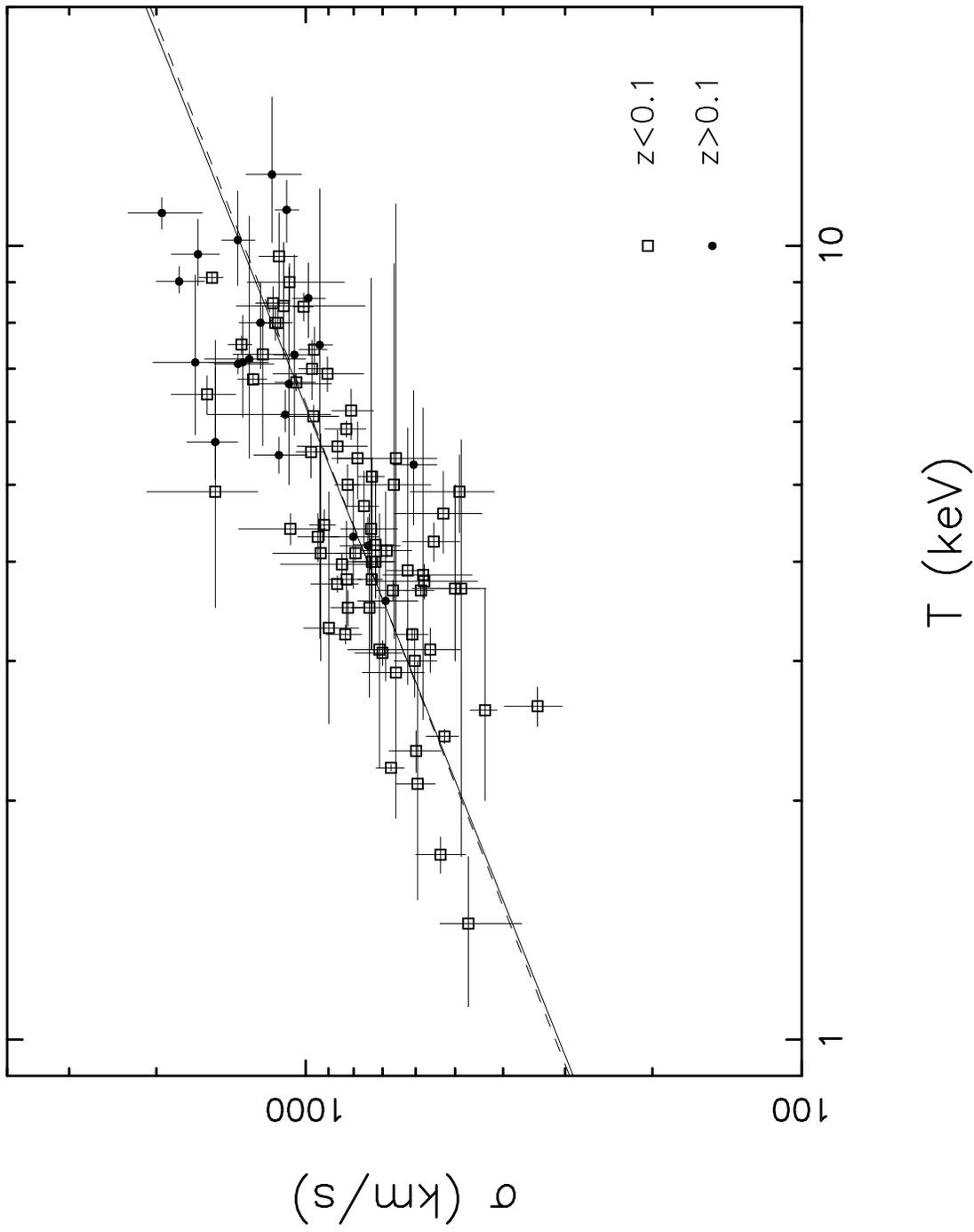
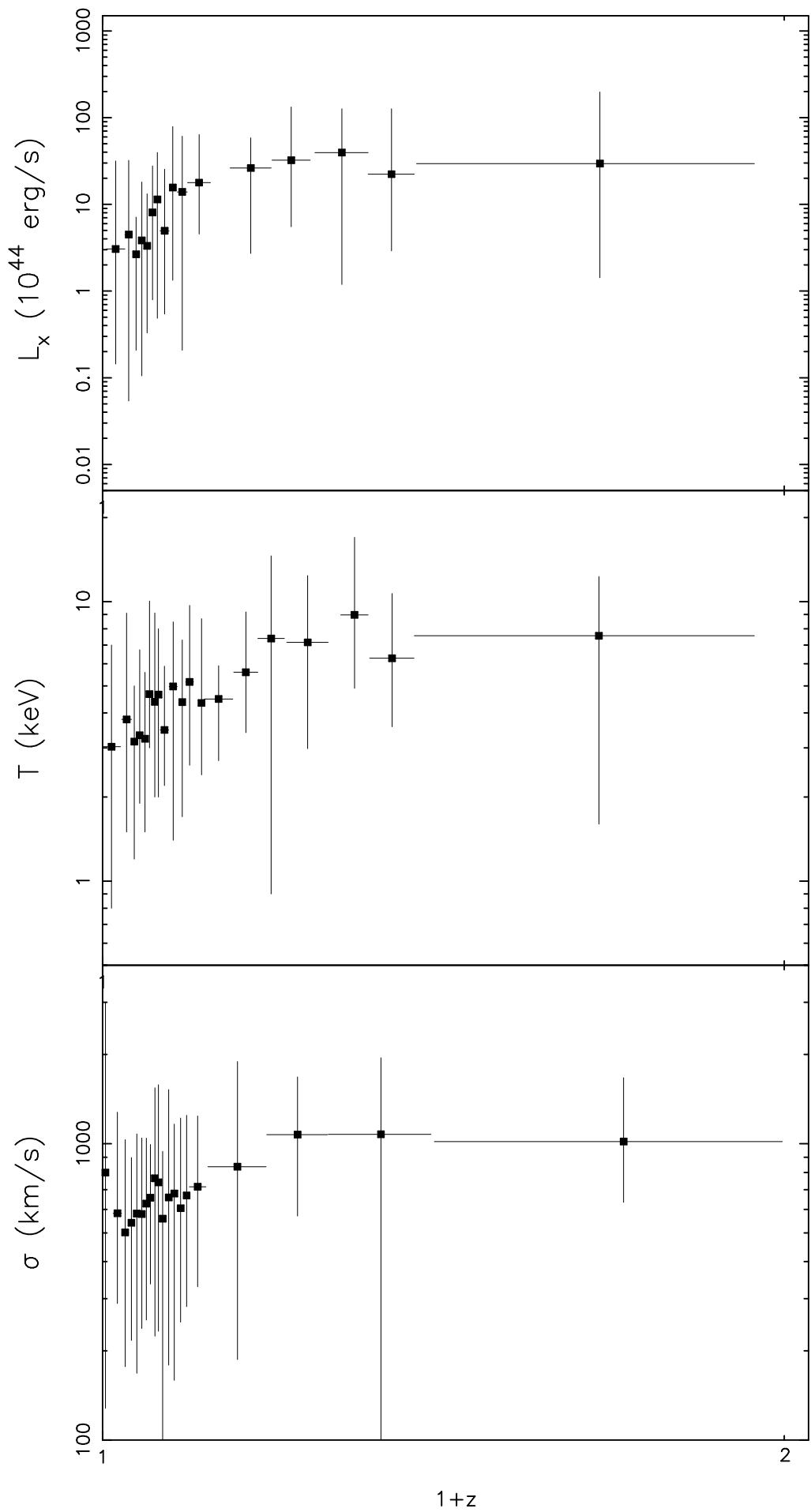
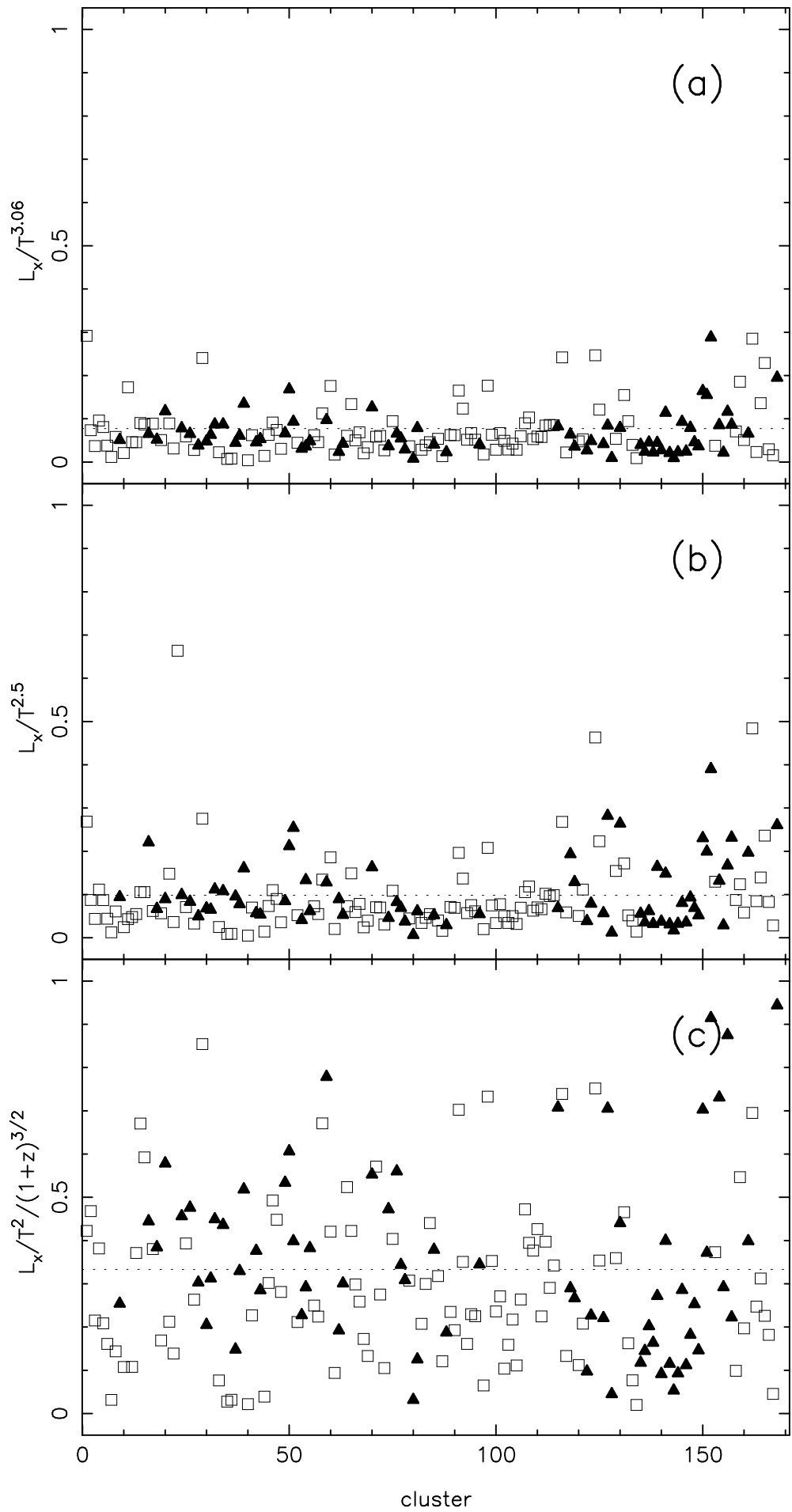
\end{document}